\documentclass[a4paper,leqno]{article}

\usepackage{amsfonts}
\usepackage{amsmath}
\usepackage{amsthm}\usepackage{amssymb}
\usepackage[mathscr]{eucal}
\usepackage{graphics}
\usepackage[pagebackref,
           colorlinks,linkcolor={blue},citecolor={red}]
           {hyperref}

\newtheorem{Theorem}{Theorem} 

\newtheorem{Lemma}{Lemma}

\newenvironment{Proof}{\noindent
\textsc{Proof --}}{\hfill$\square$\medskip}

\newcounter{remark} 
\renewcommand{\theremark}{\arabic{section}.\arabic{remark}}
\newenvironment{Remark}{
\refstepcounter{remark}
\bigskip\noindent \textit{Remark
\theremark\ - }} 

\def\bmit{\boldsymbol}

\def\u{\bmit u}
\def\w{\bmit w}
\def\h{\bmit h}
\def\e{\bmit e}
\def\a{\bmit a}

\def\f{\bmit f}

\def\x{\bmit x}
\def\n{\bmit n}

\def\w{\bmit w}
\def\z{\bmit z}

\begin{document}
\title{On the existence of  $D$--solutions of the steady--state Navier--Stokes equations\\ in 
plane  exterior domains
 }
\author{Antonio Russo 
\\{\footnotesize Dipartimento di Matematica}\\[-4pt]
{\footnotesize Seconda Universit\`a degli Studi  di Napoli}\\[-4pt]
{\footnotesize Via Vivaldi, 43}\\[-4pt]
{\footnotesize 81100 - Caserta, ITALY}\\[-4pt]
{\footnotesize   a.russo@unina2.it }
}

\date{}
\maketitle
{\small {\bf  Mathematical Subject classification (2010)}: 35Q30, 76D03, 76D05, 76D07.

\medskip

}
\medskip

\begin{abstract}
 We  prove that the steady--state  Navier--Stokes problem in a   plane Lipschitz domain $\Omega$ exterior to a bounded and simply
connected set
 has a
$D$--solution provided the boundary datum $\a \in L^2(\partial\Omega)$ satisfies
$$
{1\over 2\pi}\left|\int_{\partial\Omega}\a\cdot\n\right|<1.
$$
If $\Omega$ is of class $C^{1,1}$, we can assume $\a\in W^{-1/4,4}(\partial\Omega)$. Moreover, we show that for every
$D$--solution
$(\u,p)$ of the Navier--Stokes equations  it holds
$$\nabla p=
o(r^{-1}),\quad \nabla_k p=O(r^{\epsilon-3/2}),\quad\nabla_k\u=O(r^{\epsilon-3/4}), 
$$ for all $k\in{\Bbb N}\setminus\{1\}$ and for all
positive $\epsilon$, and if the flux of $\u$ through a circumference surrounding $\complement\Omega$ is zero, then there is a
constant vector $\u_0$ such that
$$
\u=\u_0+o(1).
$$    
\end{abstract}

\vfill\eject

\section{Introduction}
Let 
 \begin{equation}
\label{defdomest}
\Omega={\Bbb R}^2\setminus\overline{\Omega'},
\end{equation}
with $\Omega'$  bounded and  simply connected Lipschitz domain\footnote{See Remark \ref{DDMMP}.}. As is well--known,
the steady--state  Navier--Stokes problem in $\Omega$ is to find a solution
$({\u},p)$ of the  system \cite{Galdi1}\footnote{As is always possible,  we assume throughout the 
kinematical viscosity coefficient equal to 1.}
\begin{equation}
\label{NSS}
\begin{array}{r@{}l}
 \Delta{\u}- \u\cdot \nabla{\u}  & {} = \nabla p\quad  \hbox{\rm
in }\Omega,\\[4pt]
\hbox{\rm div}\,{\u} & {} =0 \! \qquad \hbox{\rm
in }\Omega,
\\[4pt]
{\u} & {} ={\a}\!\!\qquad
\hbox{\rm on }\partial\Omega,
\end{array}
\end{equation}
where $\u$, $p$ and $\a$ are respectively  the velocity, the pressure and the boundary datum. In  
\cite{arusso} we removed the classical  zero flux condition  
$$
 \int_{\partial\Omega}\a\cdot \n =0 
$$
 for the existence of a  
solution of system  (\ref{NSS})\footnote{See \cite{Galdi1}, Ch. IX, and \cite{GaldiDD}.}. Indeed, by following the  
well--known approach of  {\it invading domains} of J. Leray \cite{Ler}, we proved    existence of  a solution
$(\u_\ell,p_\ell)\in D^{1,2}(\Omega)\times L^2_{\rm loc}(\overline\Omega)$ of problem (\ref{NSS}), we shall call Leray
solution,  provided
$\a\in W^{1/2,2}(\partial\Omega)$  and 
\begin{equation}
\label{hhjyu}
{\kappa \over2\pi}\left|\int_{\partial\Omega} \a\cdot\n\right|<1,
\end{equation}
with
\begin{equation}
\label{sup}
\kappa=\sup_{\|{\bmit\varphi}\|_{D^{1,2}_\sigma({\Bbb R}^2)}=1} \left|\int_{{\Bbb
R}^2} (\log r)\hbox{\rm div}\,({\bmit\varphi}\cdot\nabla{\bmit\varphi})\right|<+\infty.
\end{equation}
By well--known results of D. Gilbarg \& H.F. Weinberger \cite{GWe}  and G.P. Galdi \cite{GaldiDD} $\u_\ell$ is
known to be   bounded in a neighborhood of infinity $\complement C_{R_0}$ and there is a (unknown)  constant vector $\u_0$ such
that
\begin{equation}
\label{PI}
\begin{array}{r@{}l}
 p_\ell(x) & {}=o(1),\\[4pt]
\u_\ell(x) & {} ={\u_0}+o(1)
\end{array}
\end{equation}
and
\begin{equation}
\label{jlkjhlkjhl}
\nabla\u_\ell=O(r^{-3/4}\log  r).
\end{equation}
Moreover, in \cite{Galdi1} it is proved that\footnote{We set
$\nabla_k\varphi = \nabla\ldots\nabla_{k-{\rm times}}\varphi$,
$\nabla_1\varphi=\nabla\varphi$, $\nabla_0\varphi= \varphi$.}
\begin{equation}
\label{P3}
\begin{array}{l}
\displaystyle
 \nabla_{k}p_\ell(x)=o(1),\\[4pt]
\displaystyle   \nabla_k \u_\ell(x)=o(1), 
\end{array}
\end{equation}
for all $k\in{\Bbb N}$.  

In a recent paper \cite{ARPJ} we improve  (\ref{P3}) by showing that 
\begin{equation}
\label{P4}
\begin{array}{l}
\displaystyle \nabla_{k}p_\ell(x)=O(r^{\epsilon-1/2}),\\[4pt]
\displaystyle    \nabla_k \u_\ell(x)=O(r^{\epsilon-1/2}),
\end{array}
\end{equation}
for all $k\in{\Bbb N}$ and for every positive $\epsilon$.
Let us note that (\ref{PI})$_1$,  (\ref{P3}) and (\ref{P4}) hold for  every solution  $(\u,p)$ of
(\ref{NSS})$_{1,2}$ such that  \cite{GW}\footnote{For a $D$--solution (\ref{jlkjhlkjhl})  is replaced by $\nabla\u
=O(r^{-3/4}\log^{9/8} r)$
\cite{GW}.}
$$
\int_{\complement C_{R_0}}|\nabla\u|^2<+\infty,
$$
for  some $C_{R_0}\Supset \Omega' $, {\it we shall call $D$--solution\/}\footnote{The existence of
a $D$--solution $(\u_f,p_f)$ can be also find by a technique of H. Fujita \cite{Fu}  (see also \cite{Galdi1}).
Due to the lack of a uniqueness theorem we cannot compare the two solutions. However, if $\u_f$ has zero outflow through
$\partial C_{R_0}$, then $\u_f$ is bounded (see Theorem \ref{Ami1}).}.   Moreover,  
 (\ref{PI})$_1$ is replaced by the weaker one
\cite{GW}
$$
|\u(x)|^2=o(\log r)
$$
If $\u$ vanishes on $\partial\Omega$, C.J. Amick proved that $\u$ is bounded so that by the results of \cite{GaldiDD}, \cite{GW}
\begin{equation}
\label{logsgsss}
\u=\u_0+o(1),
\end{equation}
 with $\u_0$ constant vector. If $\u_0\ne{\bf 0}$ L.I. Sazonov \cite{Sazonov} showed that $\u$ is
physically reasonable in the sense of R. Finn and D.R. Smith \cite{FS}, \cite{Smith} so that it behaves at infinity (almost) as
the solution of the Oseen problem\footnote{See Remark \ref{gfgfgh7}.}. To the best of our knowledge this is the state of the art
of the problem of the existence and asymptotic behavior at infinity of a
$D$--solution\footnote{For $\u_0\ne {\bf 0}$ by   different  approaches and under suitable smallness assumption on the data R.
Finn \& D.R. Smith
\cite{FS} (see also  \cite{ARSub}) and G.P. Galdi \cite{Galdi1} proved existence of a
$D$--solution of (\ref{NSS}) which takes the value $\u_0$ at infinity .}.

In this paper we continue the study started in \cite{arusso} on system (\ref{NSS}) with a threefold main purpose:

\medskip

$\bullet$  to prove that $\kappa\le1$ and to get the results of \cite{arusso} by weakening  the
hypotheses on the boundary datum; to be precise we
shall only assume $\Omega$ Lipschitz and   
$\a\in L^q(\partial\Omega)$, $q\ge 2$, and prove existence of
a $D$--solution of equations (\ref{NSS})$_{1,2}$ which takes the boundary value $\a$ 
in the sense of the nontangential convergence for $q>2$; 
if $\Omega$ is of class $C^{1,1}$, we can assume $\a\in W^{-1/4,4}(\partial\Omega)$.
 \medskip

$\bullet$ to observe that Amick's result (\ref{logsgsss}) on the boundedness of a $D$--solution  holds under the sole hypothesis
that the flux of $\u$ through $\partial C_{R_0}$ is zero; 

\medskip

$\bullet$ starting from the results of \cite{ARPJ} to show that for every $D$--solution  $(\u,p)$
\begin{equation}
\label{sharp}
\nabla p(x)=o(r^{-1})
\end{equation}
and\footnote{As will appear clear from the proof the quantity $r^\epsilon$ can be replaced by a suitable power of $\log r$ 
depending on $k$.}
$$
\begin{array}{l}
\displaystyle \nabla_{k}p(x)=O(r^{\epsilon-{3/2}}),\\[4pt]
\displaystyle   \nabla_k \u(x)=O(r^{\epsilon-3/4}),
\end{array}
$$
for all $k\in{\Bbb N}\setminus\{1\}$. Moreover,  by means of the classical Hamel solutions 
we observe that  (\ref{sharp}) is sharp.

\bigskip

{\small {\sc Notation} --  A domain (open connected set) $\Omega$ of $\,{\Bbb R}^2$ is said to be of class $C^{k,\alpha}$
 if  for every $\xi\in\partial\Omega$,
there exists a neighborhood  of $\xi$ in $\partial\Omega$ which can be expressed as a graph of a  
function of class $C^{k,\alpha}$; for $k=0$ and $\alpha=1$ $\Omega$ is said to be Lipschitz.
 We shall use a standard vector notation, as in \cite{Galdi1};
$\{o,
(\e_1,\e_2)\}$ is a cartesian reference frame of $\,{\Bbb R}^2$ with origin $o$ and $\{\e_1,\e_2\}$ orthonormal basis of $\,{\Bbb R}^2$;
$\{o,(\e_r,\e_\theta)\}$ is the polar coordinate system with origin at $o$; 
$x=(x_1,x_2)=(r,\theta)$ denotes the generic point  of $\,{\Bbb R}^2$, with $r=|\x|$, $\x=x-o=r\e_r$;  if $\u$ is a
vector field in ${\Bbb R}^2$, by
$(u_1,u_2)$ and $(u_r,u_\theta)$  we denote the cartesian and polar components of $\u$ respectively, and we set $(\nabla
\u)_{ij}= \partial u_j/\partial x_i$,
$\partial_r\u=\e_r\cdot\nabla\u$, $\partial_\theta\u=\e_\theta\cdot\nabla\u$, $\nabla^\perp=(-\partial_2, \partial_1)$.
$C_R$ is the disk of radius
$R$ centered at $o$; also, we set $T_R=C_{2R}\setminus\overline C_R$, $\Omega_R=\Omega\cap C_R$; if
$\Omega_1$ and
$\Omega_2$ are two domains,  
$\Omega_1\Subset\Omega_2$ means that $\overline{\Omega_1}\subset \Omega_2$; if $\Omega$ is the exterior
domain (\ref{defdomest}) we denote by $R_0$ a positive constant such that $  \Omega'\Subset C_{R_0}$;  the symbol
$c$ will be reserved to denote a positive constant whose numerical value is unessential to our purposes.
We use a standard notation to denote (scalar, vector or second--order tensor) function spaces, as in \cite{Galdi1} and,  in
particular
$D^{k,q}(\Omega)$ denotes the Banach space of all fields $\varphi\in L^1_{\rm loc}(\Omega)$ such that
$\|\nabla_k\varphi\|_{L^q(\Omega)}<+\infty$; $D^{k,2}_\delta(\Omega)=\{\varphi\in L^1_{\rm
loc}(\Omega):\|\sqrt\delta \nabla_k\varphi\|_{L^2(\Omega)}<+\infty\}$, where
$\delta=\delta(x)$ is a function equal to the distance of $x$ from $\partial\Omega$ in a neighborhood of
$\partial\Omega$ and to 1 in $C_{R_0}$;
${\cal H}^q$ ($q>0)$
 stands for the Hardy space  in ${\Bbb R}^2$ \cite{Stein}. The   symbol
$V_\sigma$, where $V(\subset L^1_{\rm loc}(\Omega))$ stands for the subset of 
$V$ of all  vector fields ${u}$ such that $\int_\Omega{u}\cdot\nabla\varphi =0$,  for all
$\varphi\in C^\infty_0(\Omega)$.    Let
$\varphi$ be a function in $\Omega$.  Let $\{\gamma(\xi)\}_{\xi\in\partial\Omega}$ be a family of
circular finite (not empty) triangles with vertex on $\partial\Omega$ such that
$\gamma(\xi)\setminus\{\xi\}\subset\Omega  $\footnote{As is well--known, since
$\Omega$ is Lipschitz  such a family of triangles certainly exists.};
$\varphi(x)$ is said to converge  nontangentially at the boundary if 
$$
\varphi(\xi)=\mathop{\lim_{x\to\xi}}_{(x\in\gamma(\xi))}\varphi(x)\Leftrightarrow
\varphi(x)\mathop{\longrightarrow}^{\rm nt} \varphi(\xi)
$$
for almost all $\xi\in\partial\Omega$.   
The (Landau) symbols  $f(x)=o(g(r))$ and  $f(x)=O(g(r))$  $(g>0)$   mean  respectively that
$\lim_{r\to+\infty}(f/g)=0$ and $f/g$ is bounded  in a neighborhood of infinity. If $\varphi\in L^1(\Omega)$  [or  
$\varphi\in \partial\Omega$] we use the symbol
$$
\int_\Omega\varphi\quad\left[\int_{\partial\Omega}\varphi\right]
$$
to denote the integral of $\varphi$ over $\Omega$ [on $\partial\Omega$].

}
 
\section{Some Lemmas} 
Throughout the paper we shall consider the domain   $\Omega$  defined by (\ref{defdomest}) and, as is always
possible, we assume that
$C_1\Subset\Omega'$. 

Let us start by recalling  some well--known results concerning the Stokes problem 
\begin{equation}
\label{SS}
\begin{array}{r@{}l}
\Delta{\u} & {} =\nabla p\quad  \hbox{\rm
in }\Omega,\\[4pt]
\hbox{\rm div}\,{\u} & {} =0 \;\;\;\quad  \hbox{\rm
in }\Omega,
\\[4pt]
{\u} & {} ={\a} \!\! \quad\quad  
\hbox{\rm on }\partial\Omega,
\end{array}
\end{equation}
we shall use in the sequel. 

It is well--known that if $\a\in L^2(\partial\Omega)$, then (\ref{SS}) has an analytical $D$--solution in
$\Omega$
\cite {FKV}, 
\cite{Russo}, \cite{RussoPG}
expressed by  
\begin{equation}
\label{hhhhhnb76}
{\bmit u}={\bmit v}+{\bmit\sigma},
\end{equation}
with
$$
{\bmit\sigma}(x)=-  {\e_r\over 2\pi r} \int_{\partial\Omega}\a\cdot\n  
$$
and $\n$ outward (with respect to $\Omega$) unit normal to $\partial\Omega$, such that $\u$ tends nontangentially to
$\a$ and
\begin{equation}
\label{zczbbbf}
\int_{\partial\Omega}{\bmit v}\cdot\n=0.
\end{equation}
It is unique in the class of the so--called very weak solutions \cite{aR}, \cite{RussoPG}. Moreover,  there is a constant vector
$\u_0$\footnote{${\bmit u}_0$ is determined by
$\a$ through  well--known compatibility conditions (see, $e.g.$, \cite{Galdi1}, \cite{aR}, \cite{Russo},  \cite{RussoPG}).} 
such that \cite{Mitrea},
\cite{Russo}, 
\cite{RussoPG}
\begin{equation}
\label{k˜lkjytgh7}
\begin{array}{r@{}l}
\nabla_k(\u-\u_0)=O(r^{-1-k}),\\[4pt]
\nabla_kp=O(r^{-2-k})
\end{array}
\end{equation}
and
\begin{equation}
\label{jknhhhu78}
\int_\Omega \delta(|\nabla\u|^2+|p|^2)\le c \int_{\partial\Omega}|\a|^2.
\end{equation}

Since $D^{1,2}_\delta(\Omega_R)\hookrightarrow W^{1/2,2}(\Omega_R)\hookrightarrow L^4(\Omega_R)$ 
\cite{JK},
 by (\ref{k˜lkjytgh7})$_1$ we have   in particular that $\u-\u_0\in L^4(\Omega)$.  Moreover,  it holds (see,
$e.g.$,
\cite{Mitrea}, \cite{Russo},  \cite{RussoPG}, \cite{She})
\begin{itemize} 
\item[(\i)] if $\a\in L^q(\partial\Omega)$, $q\in[2,+\infty]$, then   $\u\in W^{1/q,q}_{\rm
loc}(\overline\Omega)$.
\end{itemize}
There are two positive scalars $\mu_0(<1)$ and $\varepsilon$ depending only on $\partial\Omega$ such that 
\begin{itemize}
\item[(\i\i)]  if $\a\in C^{0,\mu}(\partial\Omega)$, $\mu\in[0,\mu_0)$, then   $\u\in C^{0,\mu}_{\rm
loc}(\overline\Omega)$; if
$\Omega$ is of class $C^1$, we  can take $\mu_0=1$;
\item[(\i\i\i)]  if $\a\in W^{1-1/q,q}(\partial\Omega)$, $q\in[2,2+\varepsilon)$, then   $(\u,p)\in W^{1,q}_{\rm
loc}(\overline\Omega)\times L^q_{\rm loc}(\overline\Omega)$; if
$\Omega$ is of class $C^1$, we  can take $q\in(1+\infty)$;
\item[(\i v)]  if $\a\in W^{1,q}(\partial\Omega)$, $q\in(2-\epsilon,2+\varepsilon)$, then   $(\u,p)\in W^{1+1/q,q}_{\rm
loc}(\overline\Omega)\times W^{1/q,q}_{\rm loc}(\overline\Omega)$; if
$\Omega$ is of class $C^1$, we  can take $q\in(1+\infty)$.  Moreover, if $\a\in W^{1,2}(\partial\Omega$, then 
$$
\int_\Omega  \big[|\nabla\u|^2+|p|^2 +\delta(|\nabla\u|^2+|p|^2)\big]\le c \int_{\partial\Omega}(|\a|^2+|\nabla\a|^2).
$$
\end{itemize}

The above results allow  us to prove
\begin{Lemma}
\label{L0}
Let $\Omega$ be a Lipschitz exterior domain of $\,{\Bbb R}^2$. If $\a\in
L^2(\partial\Omega)$, then there is a field $\h\in C^\infty_\sigma(\Omega)\cap
D^{1,2}_\delta(\Omega )$  which tends nontangentially to $\a$ on $\partial\Omega$, vanishes outside a disk and satisfies
\begin{equation}
\label{srvissima}
\|\h\|_{L^4(\Omega)}\le c\|{\bmit h}\|_{D^{1,2}_\delta(\Omega)}\le c \|\a\|_{L^2(\partial\Omega)}.
\end{equation} 
Moreover, if $\a$ is more regular, then also ${\bmit h}$ is more regular  according to $(\i)$--$(\hbox{\rm \i v})$. 
\end{Lemma}  
\begin{Proof}
Let $g$ be a $C^\infty$ cut--off function in ${\Bbb R}^2$, equal to 1 in $C_{\bar R}$
and to zero outside $C_{2{\bar R}}$ with ${\bar R}>R_0$. Since by (\ref{zczbbbf})
$$
\int_{T_{\bar R}}\hbox{\rm div}\,(g{\bmit v} )=0,
$$
the problem
$$
\hbox{\rm div}\,{\bmit\omega}+\hbox{\rm div}\,(g {\bmit v} )=0\quad\hbox{\rm in }T_{\bar R}
$$  
admits a solution ${\bmit \omega}\in C^\infty_0(T_{\bar R})$ \cite{KP1} (see also \cite{Galdi1} Ch.III). It is clear that
the field
\begin{equation}
\label{fgamma}
{\bmit h}(x)={\bmit\zeta}(x) +{\bmit\sigma},\qquad{\bmit\zeta}=\begin{cases}{\bmit
v},&\hbox{\rm in }
\Omega_{\bar R},\\
{\bmit\omega}+g{\bmit v},&\hbox{\rm in }  T_{\bar R},\\
{\bf 0},& \hbox{\rm in }{\Bbb R}^2\setminus\Omega_{2\bar R},
\end{cases}
\end{equation}
satisfies all the properties stated in the Lemma.
\end{Proof}

If $\Omega$ is of class $C^{1,1}$ and $\a\in W^{-1/q,q}(\partial\Omega)$
$(q>1)$\footnote{$W^{-1/q,q}(\partial\Omega)$ is the dual space of $W^{1-1/q',q'}(\partial\Omega)$.}, then 
(\ref{SS}) admits the solution (\ref{hhhhhnb76}) where
${\bmit v}$ is a simple layer potential (plus a constant vector
$\u_0$)  with a density in $W^{-1-1/q,q}(\partial\Omega)$
\cite{DL}, \cite{RussoPG}. The boundary  datum is taken in the sense of the unique continuous extension map  from
$W^{-1-1/q,q}(\partial\Omega)$ into $W^{ -1/q,q}(\partial\Omega)$ of the trace operator of the classical simple layer potential
 from $W^{1/q,q}(\partial\Omega)$ to $W^{1+1/q,q}(\partial\Omega)$ \cite{DL} . Moreover, $\u\in L^q_{\rm
loc}(\overline\Omega)$ satisfies (\ref{k˜lkjytgh7}).  Therefore, by proceeding as we did in the proof of Lemma \ref{L0} and taking also
into account the regularity properties of the classical   layer potentials \cite{Mi2}, \cite{RussoPG}, we have 
\begin{Lemma}
\label{L01}
Let $\Omega$ be a  exterior domain of $\,{\Bbb R}^2$ of class $C^{1,1}$. If $\a\in
W^{-1/q,q}(\partial\Omega)$, $q\ge 4$,  then there is  a divergence free extension   $\h\in
C^\infty_\sigma(\Omega)\cap L^q( \Omega)$ of $\a$ in $\Omega$, expressed by $(\ref{fgamma})$, which
satisfies 
\begin{equation}
\label{srvissima1}
\|\h\|_{L^q(\Omega)}\le c \|\a\|_{W^{-1/q,q}(\partial\Omega)}.
\end{equation} Moreover, 
\begin{itemize}
\item if $\a\in C^{1,\mu}(\partial\Omega)$, $\mu<1$, then  $\h\in C^{1,\mu}(\overline\Omega)$;
\item if $\Omega$ is of class $C^k$, $k\ge 2$ and $\a\in W^{k-1/q,q}(\partial\Omega)$, $\mu<1$, then  $\h\in
W^{k+2,q}(\Omega) $.
\end{itemize}

\end{Lemma}

The following elementary but basic Lemma was first proved in \cite{GW}. We give a simple proof of a slight 
generalization.
\begin{Lemma}
\label{L1}
Let $\w, {\bmit z}\in W^{1,2}_\sigma(C_R)$. Then
\begin{equation}
\label{GWtt}
 \left|\int_{C_R} \w\cdot\nabla\z\cdot{\e_r\over r}\right|\le  \left\{\int_{C_R}|\nabla
\w|^2 \int_{C_R}|\nabla\z|^2\right\}^{1/2}.
\end{equation}
\end{Lemma} 
\begin{Proof}
Set
$$
\bar\varphi(r)={1\over2\pi}\int_0^{2\pi}\varphi(r,\theta).
$$
Since
$$
\int_0^{2\pi}\bar w_1  \partial_\theta z_2 =\int_0^{2\pi}\bar w_2   \partial_\theta z_1=0  ,
$$
a simple computation yields \cite{aR}
\begin{equation}
\label{poaf}
 \int_{C_R }
\w\cdot\nabla\z\cdot{\e_r\over r} =\int_0^R{1\over\rho}\int_0^{2\pi}\big[(
 w_1-\bar w_1) \partial_\theta z_2 
-(w_2 -\bar w_2) \partial_\theta z_1\big].
\end{equation}
From Schwarz's, Wirtinger's and Cauchy's inequalities  we have
$$
\begin{array}{r@{}l}
\displaystyle\left|\int_0^{2\pi} (
 w_1-\bar w_1) \partial_\theta z_2 \right| & {}\displaystyle \le\left\{\int_0^{2\pi}|w_1-\bar
w_1|^2\int_0^{2\pi}|\partial_\theta z_2|^2\right\}^{1/2}\\[12pt]
 & {}\displaystyle \le \left\{\int_0^{2\pi}|\partial_\theta w_1|^2\int_0^{2\pi}|\partial_\theta
z_2|^2\right\}^{1/2} ,\\[12pt]
\displaystyle\left|\int_0^{2\pi} (
 w_2-\bar w_2) \partial_\theta z_1 \right| & {}\displaystyle \le \left\{\int_0^{2\pi}|\partial_\theta
w_2|^2 \int_0^{2\pi}|\partial_\theta z_1|^2\right\}^{1/2}.
\end{array} 
$$
Therefore, (\ref{GWtt}) follows from (\ref{poaf}),  taking into account that $|\partial_\theta\w|\le
r|\nabla
\w|$. 
\end{Proof}

Note that if $\w, \;\z\in D^{1,2}_{\sigma}({\Bbb R}^2)$, letting $R\to+\infty$ in (\ref{GWtt}) yields
\begin{equation}
\label{GWaaa}
 \left|\int_{{\Bbb R}^2} \w\cdot\nabla\z\cdot{\e_r\over r}\right|\le \left\{ \int_{{\Bbb R}^2}|\nabla
\w|^2 \int_{{\Bbb R}^2}|\nabla\z|^2\right\}^{1/2}.
\end{equation}
If $\w,\z\in W^{1,2}_{\sigma, 0}(\Omega)$, then the zero extensions of $\w$ and $\z$ belongs to $D^{1,2}_\sigma({\Bbb
R}^2)$. Therefore,  from (\ref{GWaaa}) it follows 
\begin{equation}
\label{GWtccc}
 \left|\int_{\Omega} \w\cdot\nabla\z\cdot{\e_r\over r}\right|\le  \left\{\int_{\Omega}|\nabla
\w|^2 \int_{\Omega}|\nabla\z|^2\right\}^{1/2}.
\end{equation}

Lemma \ref{L1} allows to quickly prove 
\begin{Theorem} It holds
\begin{equation}
\label{iuytoiuy}
\sup_{\|{\bmit\varphi}\|_{D^{1,2}_\sigma({\Bbb R}^2)}=1} \left|\int_{{\Bbb
R}^2} (\log|\x|)\hbox{\rm div}\,({\bmit\varphi}\cdot\nabla{\bmit\varphi})\right|\le1.
\end{equation}
\end{Theorem}
\begin{Proof} It is sufficient to prove (\ref{iuytoiuy}) in $C^\infty_{\sigma,0}({\Bbb R}^2)$.  In this case
$$ 
\left|\int_{{\Bbb
R}^2} (\log|\x|)\hbox{\rm div}\,({\bmit\varphi}\cdot\nabla{\bmit\varphi})\right|=\left|\int_{{\Bbb
R}^2}{\bmit\varphi}\cdot\nabla{\bmit\varphi}\cdot{\e_r\over r} \right|
$$
and (\ref{iuytoiuy}) follows from Lemma \ref{L1}.
\end{Proof}

\begin{Lemma}
\label{funlrlele}
 Let $(\u,p)$ be a a solution to  $(\ref{NSS})_{1,2}$. Then  for all $k\in{\Bbb N}_0$ and for all 
$C_1(x)\Subset\Omega$
\begin{equation}
\label{otyu}
|\nabla_kp(x)|\le c\left\{\int_0^{2\pi}|\nabla_k p|(|\x|,\theta)
+\sum_{j=1}^{k+1}\|\nabla_j\u\|_{L^2(C_1(x))}^2\right\}.
\end{equation}
\end{Lemma}
\begin{Proof} We follow \cite{GaldiDD} (Lemma 3.10).  Setting $\u_1=\nabla \u$ and $p_1=\nabla p$, the pair  $(\u_1,p_1)$ is a
solution of the equations
\begin{equation}
\label{NSS1}
\begin{array}{r@{}l}
\displaystyle \Delta{\u}_1-  \u_1\cdot \nabla{\u}-\u \cdot \nabla{\u}_1  & {} = \nabla p_1 ,\\[4pt]
\hbox{\rm div}\,{\u}_1 & {} =0.
\end{array}
\end{equation}
Let $x=(r,\theta)$ and let $(r',\theta')$ be a polar coordinate system centered at $x$. 
Multiplying (\ref{NSS1})$_1$ scalarly by $\x'/{r'}^2 $ and integrating over $C_1(x)$, we have
$$
p_1(r,\theta)= p_1(r',\theta') +\int_{C_1(x)}[\u_1\cdot\nabla\u+\u\cdot\nabla\u_1]\cdot{\x' \over{r'}^2 }
$$
Hence, making use of Lemma \ref{L1},  it follows
\begin{equation}
\label{gfdjh8}
\begin{array}{ l}
|p_1(r,\theta)|  \displaystyle\le {1\over
2\pi} \left\{\int_0^{2\pi}|p_1(r',\theta')|+\|\nabla\u\|_{L^2(C_1(x))}\|\nabla\u_1\|_{L^2(C_1(x))}
\right\}\\[10pt]
 \displaystyle \int_0^{2\pi}|p_1(r',\theta') | \le 2\pi
\left\{|p_1(r,\theta)|+\|\nabla\u\|_{L^2(C_1(x))}\|\nabla\u_1\|_{L^2(C_1(x))}.
\right\}
\end{array}
\end{equation}
Multiplying (\ref{gfdjh8})$_2$ by $r'$ and integrating over $r'\in[0,1]$ and $\theta\in[0, 2\pi]$ show that
 \begin{equation}
\label{kjh˜ljkh8}
\|p_1\|_{L^1(C_1(x))}\le
c\left\{\int_0^{2\pi}|p_1(r ,\theta )|+\|\nabla\u\|_{L^2(C_1(x))}\|\nabla\u_1\|_{L^2(C_1(x))}\right\}.
\end{equation}
Moreover, multiplying (\ref{gfdjh8})$_1$ by $r'$ and integrating over $r'\in[0,1]$ and $\theta'\in[0,2\pi]$ yield
 \begin{equation}
\label{kjh˜ljkh9}
 |p_1(x)|\le
c\left\{\|p_1\|_{L^1(C_1(x))}+\|\nabla\u\|_{L^2(C_1(x))}\|\nabla\u_1\|_{L^2(C_1(x))}\right\}.
\end{equation}
Therefore, putting together (\ref{kjh˜ljkh8})--(\ref{kjh˜ljkh9}) and using Cauchy's inequality we find  
$$
\begin{array}{r@{}l}
 |p_1(x)| & {} \displaystyle\le
c\left\{\int_0^{2\pi}|p_1(r ,\theta )|+\|\nabla\u\|_{L^2(C_1(x))}\|\nabla\u_1\|_{L^2(C_1(x))}\right\}\\[10pt]
 & {} \displaystyle 	\le c\left\{\int_0^{2\pi}|p_1(r ,\theta )|+\|\nabla\u\|_{L^2(C_1(x))}^2+\|\nabla\u_1\|_{L^2(C_1(x))}^2\right\}.
\end{array}
$$
and (\ref{otyu}) is proved for $k=1$.  The proof for $k=0$ follows the same steps. Iterating such a procedure as many times
as we need, we then prove (\ref{otyu}).
\end{Proof}

\begin{Lemma}
 \label{lopimjk} {\rm \cite{Galdi1}} Let
$$
v(x)=\int_{{\Bbb R}^2}{1\over |x-y|^\lambda|y|^\mu}{\rm d}a_y,
$$
with $\lambda<2$, $\mu<2$. If  $\lambda+\mu>2$, then
$$
v(x)= cr^{2-\lambda-\mu},
$$
for a suitable constant $c=c(\lambda,\mu)$.
\end{Lemma}

\begin{Lemma} {\rm \cite{Stein}}
\label{lopimjk1}
If $f\in {\cal H}^1 $, then the problem
$$
\begin{array}{r@{}l}
\Delta p & {} =f\quad\hbox{\rm in }{\Bbb R}^2,\\[4pt]
\displaystyle\lim_{x\to\infty}p(x)&{}=0
\end{array}
$$
admits the unique solution
$$
p(x)={1\over 2\pi}\int_{{\Bbb R}^2}f(y)\log|x-y|{\rm d}a_y\in D^{2,1}({\Bbb R}^2)\cap
D^{1,2}({\Bbb R}^2).
$$
 
\end{Lemma}
\begin{Lemma} {\rm \cite{CLMS}}
\label{lopimjk2}
 If $\u \in D^{1,2}_\sigma({\Bbb R}^2)$, then
$\nabla\u\cdot\nabla\u^{\rm T}\in{\cal H}^1 $.
\end{Lemma} 

\section{Asymptotic behavior of $D$--solutions}

Let us recall that  by $D$--solution we mean an analytical  pair $(\u,p)$ which satisfies equations
$(\ref{NSS})_{1,2}$ and 
\begin{equation}
\label{hgjytt7}
\int_{\complement C_{R_0}}|\nabla\u|^2<+\infty,
\end{equation}
for some $C_{R_0}\Supset\Omega'$. 
 
We deal now with  the asymptotic properties of a $D$--solution. 
To this end we need the following classical  results  of D. Gilbarg and H.F. Weinberger \cite{GW}.
\begin{Lemma}
\label{GWLLL}
If $(\u,p)$ is a D--solution, then
\begin{equation}
\label{hgkjhgjgh0}
\lim_{x\to+\infty} p(x)=0 
\end{equation} 
and
\begin{equation}
\label{jhljhu76}
\u=o(\sqrt{\log r}).
\end{equation}
Moreover
\begin{equation}
\label{lkj˜jk12}
\nabla\u =O(r^{-3/4}\log^{9/8} r).
\end{equation}
\end{Lemma}
Also, it holds  \cite{ARPJ}
\begin{Lemma}
\label{ARPPP}
If $(\u,p)$ is a D--solution, then  
\begin{equation}
\label{P422w}
\begin{array}{l}
\displaystyle \nabla_{k}p (x)=O(r^{\epsilon-1/2}),\\[4pt]
\displaystyle    \nabla_k \u (x)=O(r^{\epsilon-1/2}),
\end{array}
\end{equation}
for all $k\in{\Bbb N}$.
\end{Lemma}
The following theorem  extends to more general boundary data a classical result of C.J. Amick \cite{Amick2}, D. Gilbarg and H.F.
Weinberger
\cite{GW} and G.P. Galdi \cite{GaldiDD}.
\begin{Theorem}
\label{Ami1}
Let  $(\u,p)$ be a D--solution. If
\begin{equation}
\label{ioupou7}
\int_{\partial C_{R_0}} u_{R_0}=0,
\end{equation}
then there is a constant vector $\u_0$ such that
\begin{equation}
\label{kjsskh6}
\u=\u_0+o(1).
\end{equation}
\end{Theorem}
\begin{Proof}
From (\ref{NSS})$_2$ and (\ref{ioupou7}) it follows that there is a  regular function $\psi$ such
that $\u=\nabla^\perp\psi$. Therefore, we can repeat the argument of Section 2.1 of \cite{Amick2} to see that there is a curve
connecting a point of $\partial C_{R_0}$ to infinity along which  the Bernoulli function $\Phi=p+{1\over 2}|\u|^2$ is monotone
decreasing ((b) of Theorem 11) and this is sufficient to assert that $\u$ is bounded (Theorem 12). Hence by Theorem 4 of
\cite{GW} there is a constant vector $\u_0$ such that
\begin{equation}
\label{jkiu87}
\lim_{r\to+\infty}\int_0^{2\pi}|\u(r,\theta)-\u_0|^2=0.
\end{equation}
Since $\nabla\u\in L^q(\complement C_{R_0})$ for all $q\ge2 $ (see, $e.g.$, \cite{Galdi1} Lemma X.3.2), (\ref{jkiu87}) implies
(\ref{kjsskh6}) by virtue of Lemma 3.10 of \cite{GaldiDD}.
\end{Proof}
 
The following theorem concerns the asymptotic behavior of the derivatives of a $D$--solution.
\begin{Theorem} 
\label{lkoinb}
If $(\u,p)$ is a D--solution, then
\begin{equation}
\label{hg2jgh01}
 p(x)\in D^{1,2}(\complement C_{R_0}), 
\end{equation}
\begin{equation}
\label{hg2jgh0}
\nabla p(x)=o(r^{-1})
\end{equation}
and
\begin{equation}
\label{hg2jgh2}
\begin{array}{r@{}l}
\nabla_{k+1} p(x) & {}=O(r^{\epsilon-3/2}),\\[4pt]
\nabla_k\u & {} =O(r^{\epsilon-3/4}),
\end{array}
\end{equation}
for every positive $\epsilon$  and for every $k\in{\Bbb N}$. Moreover, if $\u$ satisfies $(\ref{ioupou7})$, then $p\in D^{1,2}(\complement
C_{R_0})$.
\end{Theorem}

\medskip

{\sc Proof of } (\ref{hg2jgh01}), (\ref{hg2jgh0}).

\medskip

Let  
$$
{\bmit\gamma}(x)= {\e_r\over r}\int_{\partial C_{R_0}} u_{R_0}
$$ 
and set
$$
\u={\bmit v}+{\bmit\gamma}.
$$
Let $g$ be a regular cut--off function in ${\Bbb R}^2$, vanishing in $C_{\bar R}$ and equal to 1
outside
$C_{2{\bar R}}$, with ${\bar R}\gg R_0$. Since
$$
\int_{\partial C_{\bar R}}{\bmit v}\cdot\n=0,
$$
the problem
$$
\hbox{\rm div}\,{\bmit h}+\hbox{\rm div}\,(g{\bmit v})=0\quad\hbox{\rm in }T_{\bar R}
$$
has a solution ${{\bmit h}}\in C^\infty_0(T_{\bar R})$ \cite{Galdi1}.
From  (\ref{NSS})$_{1,2}$ it follows that  the function $Q=g^2 p$ is a solution of the equation
\begin{equation}
\label{POEQ1}
\Delta Q +\sum_{i=1}^3\hbox{\rm div}\,{{\bmit h}}_i+\varphi=0\quad \hbox{\rm in } {\Bbb R}^2,
\end{equation}
where
$$
\varphi\in C^\infty_0(T_{\bar R})
$$ 
and
$$
\begin{array}{l}
{{\bmit h}}_1=(g{\bmit v}+{{\bmit h}})\cdot\nabla(g{\bmit v}+{{\bmit h}}),\\[4pt]
{{\bmit h}}_2=2 g{\bmit v} \cdot\nabla(g{\bmit\gamma}),\\[4pt]
{{\bmit h}}_3=  g{\bmit\gamma} \cdot\nabla(g{\bmit\gamma}).
\end{array}
$$
By virtue of (\ref{hgkjhgjgh0}) equation (\ref{POEQ1}) has a  unique solution   $Q$ which 
by Lemma \ref{lopimjk} is expressed by
\begin{equation}
\label{POEQ4}
\begin{array}{r@{}l}
2\pi Q(x) &{}\displaystyle=-\sum_{i=1}^3\int_{{\Bbb R}^2}(\log|x-y|)\hbox{\rm div}\,{{\bmit h}}_i(y){\rm
d}a_y-\int_{{\Bbb R}^2}\varphi(y)\log|x-y |{\rm d}a_y\\[12pt]
&{}\displaystyle=\sum_{1=1}^4Q_i.
\end{array} 
\end{equation}
By Lemma \ref{lopimjk2}  $\hbox{\rm div}\,{{\bmit h}}_1\in {\cal H}^1 $ so that Lemma
\ref{lopimjk1} implies that $Q_1\in D^{2,1}({\Bbb R}^2)\cap D^{1,2}({\Bbb R}^2)$. Hence it follows in particular that
\begin{equation}
\label{oiu1}
\lim_{x\to+\infty} Q_1(x) \in{\Bbb R}.
\end{equation}

 Now, letting
$R\to+\infty$ in the relation
$$
\begin{array}{r@{}l}
\displaystyle \int_{C_R} (\log|x-y|)\hbox{\rm div}\,[(g{\bmit v})\cdot\nabla (g{\bmit\gamma})]{\rm
d}a_y &{}\displaystyle=\int_{\partial C_R}(\log|x-\zeta|) ({\bmit v} \cdot\nabla  {\bmit\gamma}\cdot\e_R)(\zeta){\rm
d}s_\zeta\\[12pt]
&{}\displaystyle+\int_{C_R} {[g{\bmit v} \cdot\nabla(g  {\bmit\gamma})](y)\cdot(x-y)
\over|x-y|^2}{\rm d}a_y,\\[12pt]
\displaystyle \int_{C_R} (\log|x-y|)\hbox{\rm div}\,[(g{\bmit\gamma})\cdot\nabla (g{\bmit\gamma})]{\rm
d}a_y &{}\displaystyle=\int_{\partial C_R}(\log|x-\zeta|) ({\bmit\gamma} \cdot\nabla 
{\bmit\gamma}\cdot\e_R)(\zeta){\rm d}s_\zeta\\[12pt]
&{}\displaystyle+\int_{C_R}{[g{\bmit\gamma} \cdot\nabla(g  {\bmit\gamma})](y)\cdot(x-y){\rm d}a_y
\over|x-y|^2}
\end{array}
$$ 
and taking into account the  behavior at infinity of ${\bmit v}$ and ${\bmit\gamma}$,
we have
\begin{equation}
\label{ghyvcvfr56}
\begin{array}{r@{}l}
Q_2(x) &{}\displaystyle=-2\int_{{\Bbb R}^2} {[g{\bmit v} \cdot\nabla(g  {\bmit\gamma})](y)\cdot(x-y)
\over|x-y|^2}{\rm d}a_y, \\[12pt]
Q_3(x) &{}\displaystyle=-\int_{{\Bbb R}^2} {[g{\bmit\gamma} \cdot\nabla(g 
{\bmit\gamma})](y)\cdot(x-y),
\over|x-y|^2}{\rm d}a_y.
\end{array} 
\end{equation}

Now, for large $|\x|$
$$
\begin{array}{r@{}l}
\displaystyle\int_{{\Bbb R}^2} {[g{\bmit v} \cdot\nabla(g  {\bmit\gamma})](y)\cdot(x-y)
\over|x-y|^2}{\rm d}a_y & {} \displaystyle= \int_{{\Bbb R}^2\setminus C_1(x)} {[g{\bmit v} \cdot\nabla(g 
{\bmit\gamma})](y)\cdot(x-y)
\over|x-y|^2}{\rm d}a_y  \\[12pt]
&{}\displaystyle + \int_{ C_1(x)} {({\bmit v}\cdot\nabla{\bmit\gamma})(y)\cdot(x-y)
\over|x-y|^2}{\rm d}a_y  \\[12pt]
 & {} \displaystyle= \int_{{\Bbb R}^2\setminus C_1(x)} {[g{\bmit v} \cdot\nabla(g 
{\bmit\gamma})](y)\cdot(x-y)
\over|x-y|^2}{\rm d}a_y  \\[12pt]
&{}\displaystyle - \int_{ \partial C_1(x)}(\log|x-y|)({\bmit v}\cdot\nabla{\bmit\gamma})(\zeta)\cdot \n(\zeta) {\rm
d}a_\zeta 
\\[12pt]
  &{}\displaystyle + \int_{  C_1(x)}(\log|x-y|)(\nabla{\bmit v}\cdot\nabla{\bmit\gamma}^{\rm T})(y){\rm d}a_y.  
\end{array}
$$ 
Hence
\begin{equation}
\label{kjiu87ui}
\begin{array}{r@{}l}
\nabla\displaystyle Q_2(x) & {} \displaystyle= -2\nabla \int_{{\Bbb R}^2\setminus C_1(x)} {[g{\bmit v} \cdot\nabla(g 
{\bmit\gamma})](y)\cdot(x-y)
\over|x-y|^2}{\rm d}a_y \\[12pt]
&{}\displaystyle +2\int_{ \partial C_1(x)}{(x-y)({\bmit v}\cdot\nabla{\bmit\gamma})(\zeta)\cdot \n(\zeta)\over
|x-y|^2 }{\rm d}a_\zeta 
\\[12pt]
  &{}\displaystyle -2\int_{         C_1(x)}{( x-y ) (\nabla{\bmit v}\cdot\nabla{\bmit\gamma}^{\rm T})(y)\over |x-y|^2}{\rm
d}a_y.
\end{array}
\end{equation}
Likewise, since
\begin{equation}
\label{kjiu87ui2}
\begin{array}{r@{}l}
\nabla\displaystyle Q_3(x) & {} \displaystyle=- 2\nabla \int_{{\Bbb R}^2\setminus C_1(x)} {[g{\bmit\gamma}
\cdot\nabla(g  {\bmit\gamma})](y)\cdot(x-y)
\over|x-y|^2}{\rm d}a_y \\[12pt]
&{}\displaystyle +2\int_{ \partial C_1(x)}{(x-y)({\bmit\gamma}\cdot\nabla{\bmit\gamma})(\zeta)\cdot \n(\zeta)\over
|x-y|^2 }{\rm d}a_\zeta 
\\[12pt]
  &{}\displaystyle -2\int_{C_1(x)}{( x-y )(\nabla{\bmit\gamma}\cdot\nabla{\bmit\gamma}^{\rm T})(y)\over |x-y|^2}{\rm d}a_y,
\end{array}
\end{equation}
taking into account the asymptotic properties of ${\bmit v}$, $\nabla{\bmit v}$ and ${\bmit\gamma}$,
(\ref{ghyvcvfr56}), (\ref{kjiu87ui}) and (\ref{kjiu87ui2}) imply
\begin{equation}
\label{oiu2}
Q_2(x),\;  Q_3(x)=O(r^{\epsilon-1}) 
\end{equation}
 and
\begin{equation}
\label{forsrsr0}
\nabla Q_2(x),\;\nabla Q_3(x)=O(r^{\epsilon-2}),
\end{equation}
for all positive $\epsilon$. By virtue of (\ref{hgkjhgjgh0})\footnote{Otherwise $p(x)$ behaves at infinity as $\log r$.}
$$
\int_{{\Bbb R}^2}\varphi=0
$$
so that 
\begin{equation}
\label{forsrsr02}
 \nabla_k Q_4(x)=O(r^{-1-k}),
\end{equation}
for all $k\in{\Bbb N}$, and (\ref{hg2jgh01}) is proved. By the basic calculus and (\ref{P3})$_1$
$$
\int_0^{2\pi}|\nabla Q_1|(R,\theta)=\left|\int_R^{+\infty}\partial_r\nabla Q_1\right|\le {c\over R}\int_{\complement S_R}|\nabla_2
Q_1|.
$$
Hence
\begin{equation}
\label{forwwrsr02}
\lim_{R\to+\infty}\left\{R\int_0^{2\pi}|\nabla Q_1|(R,\theta)\right\}=0.
\end{equation}
Then, (\ref{hg2jgh0}) follows from Lemma \ref{funlrlele}, taking into account  (\ref{forsrsr0}),
(\ref{forsrsr02}),
 (\ref{forwwrsr02})  and that $p(x)=Q(x)$
for large $|\x|$.

 \bigskip

{\sc Proof of} (\ref{hg2jgh2}).

\medskip

Since
$$
\Delta p=-\nabla\u\cdot\nabla\u^{\rm T}\quad\hbox{\rm in }\complement C_{R_0},
$$
writing the Stokes formula in $S_R\cap\complement C_{R_0}$, taking the gradient, letting $R\to+\infty$ and taking into account
(\ref{hgkjhgjgh0}), (\ref{P422w})$_1$,  we have
\begin{equation}
\label{StPh}
\begin{array}{r@{}l}
 2\pi\nabla p(x) & {} \displaystyle=- \int_{\partial C_{R_0}}{(x-\zeta)\partial_r
p(\zeta)\over|x-\zeta|^2}{\rm d} s_\zeta+\nabla\int_{\partial C_{R_0}}{p(\zeta)(x-\zeta)\cdot
\e_{R_0}\over|x-\zeta|^2} {\rm d} s_\zeta\\[12pt] & {}
\displaystyle-\int_{\complement C_{R_0}}{(x-y)(\nabla\u\cdot\nabla\u^{\rm T})(y)\over|x-y|^2} {\rm
d}a_y \\[12pt] & {}
\displaystyle=-\int_{\complement C_{R_0}}{(x-y)(\nabla\u\cdot\nabla\u^{\rm T})(y)\over|x-y|^2} {\rm
d}a_y +\psi(x) \\[12pt]
& {} \displaystyle=-\int_{ 
\complement C_{R_0}\setminus C_1(x)}{(x-y)(\nabla\u\cdot\nabla\u^{\rm T})(y)\over|x-y|^2} {\rm
d}a_y 
\\[12pt] & {} \displaystyle+\int_{ 
 \partial C_1(x)}  (\log|x-\zeta|)(\nabla\u\cdot\nabla\u^{\rm T})(\zeta)\n(\zeta)   {\rm
d}s_\zeta \\[12pt]
 & {} \displaystyle-\int_{ 
  C_1(x)}{ (\log|x-y|)\nabla (\nabla\u\cdot\nabla\u^{\rm T}) (y)  } {\rm
d}a_y+\psi(x),
\end{array}
\end{equation}
with
$$
\nabla_k\psi(x)=O(r^{-1-k}).
$$
Hence, taking the gradient, it follows
\begin{equation}
\label{jjjfgt}
2\pi\nabla_2 p(x)  =\sum_{i=1}^3{\cal J}_i+O(r^{-2}), 
\end{equation}
where
$$
\begin{array}{ l}
\displaystyle{\cal J}_1=-\nabla\int_{ 
\complement C_{R_0}\setminus C_1(x)} {(\nabla\u\cdot\nabla\u^{\rm T})(y)(x-y)\over|x-y|^2}  {\rm
d}a_y,
\\[12pt]
\displaystyle{\cal J}_2=-\int_{ 
  C_1(x)}{ (x-y)\otimes\nabla (\nabla\u\cdot\nabla\u^{\rm T}) (y)\over|x-y|^2  } {\rm
d}a_y,\\[12pt]
\displaystyle{\cal J}_3=   \nabla\int_{\partial C_1(x)}  (\log|x-\zeta|)(\nabla\u\cdot\nabla\u^{\rm
T})(\zeta)(x-\zeta)\n(\zeta) {\rm d}s_\zeta .
\end{array}
$$
Setting $|\x|=R$ $(R>R_0)$ and after portioning $\complement C_{R_0}$ into $\complement C_{R_0}\cap C_{R/2}$, $\complement
C_{R/2}$ and taking into account (\ref{lkj˜jk12}), we get
$$
\begin{array}{ l}
\displaystyle|{\cal J}_1(x)|\le {c\over R^2}\int_{\complement C_{R_0}\cap
C_R}|\nabla\u|^2+c\int_{\complement C_{R}} r^{\epsilon-7/2}\le c|{\bmit x}|^{\epsilon -3/2}
\\[12pt]
\end{array}
$$
for some positive constant $c$ independent of $R$. Also, by (\ref{lkj˜jk12})--(\ref{P422w}) it is readily seen that  $ {\cal
J}_2(x),{\cal J}_3(x)=O(r^{\epsilon -3/2})$. Hence (\ref{hg2jgh2}) follows for $k=2$.
The
proof of $(\ref{hg2jgh2})_1$ for general $k$ is obtained by iterating  the above
argument.

The proof of $(\ref{hg2jgh2})_2$ follows the above steps, taking into account that by (\ref{jhljhu76}),
(\ref{lkj˜jk12}) and (\ref{hg2jgh0})
$$
\Delta\u  = \u\cdot\nabla\u+\nabla p =O(r^{-3/4}\log^{13/8} r).\eqno\square
$$

\begin{Remark}
Note that under assumption (\ref{ioupou7})
 in (\ref{POEQ4}) $Q_2=Q_3=0$ so that  $p\in D^{2,1}(\complement C_{R_0})$ \cite{ARPJ}.\hfill$\diamond$
\end{Remark}

\begin{Remark}
 The first basic result in \cite{GW} assures that  
\begin{equation}
\label{j˜lkj˜kjn0}
\nabla_2\u\in L^2(\complement C_{R_0}). 
\end{equation}
Hence, taking into account  
(\ref{jhljhu76}) and (\ref{NSS})$_1$, it follows that $\nabla p/\sqrt{\log r}\in L^2(\complement C_{R_0})$.  This is sufficient to
say that   (\ref{POEQ1}) has the  unique solution  (\ref{POEQ4}). Therefore, (\ref{hgkjhgjgh0}) follows from (\ref{oiu1}),
(\ref{oiu2}) and (\ref{forsrsr02}). In this way we gave an alternative proof of (\ref{hgkjhgjgh0}) based only on  (\ref{j˜lkj˜kjn0}). Note
that from (\ref{j˜lkj˜kjn0}) and  (\ref{hg2jgh0}) it follows that  
$
\u\cdot\nabla\u \in L^2(\complement C_{R_0}).$\hfill$\diamond$
\end{Remark}

 \begin{Remark}
\label{gfgfgh7}
The   asymptotic results in Theorem \ref{lkoinb} are new in the case where $\u$ is unbounded\footnote{By Theorem \ref{Ami1} this
could happens only if $\int_{\partial C_{R_0}}u_r\ne 0$.} or tends to zero at large distance. Indeed, if $\u$ tends to $\e_1$
(say)  at infinity,   L.I. Sazonov
\cite{Sazonov} showed that
$(\u,p)$ is physically meaningful in the sense of R. Finn an D.R. Smith
 \cite{FS}, \cite{Smith}. Therefore the solution enjoys the following  summability properties 
(see, $e.g.$,  \cite{Galdi1} Ch.  X) 
\begin{equation}
\label{Poklmj}
\begin{array}{l}
u_1-1\in L^q(\Omega),\;u_2\in L^{q-1}(\Omega),\;\: p\in L^{q-1}(\complement C_{R_0}),\quad\forall\,q>3,\\[4pt]
\partial_2 u_1\in
L^s(\complement C_{R_0}),\quad\forall\,s>3/2,\\[4pt]
\partial_1 u_1,\nabla u_2,\; \nabla_2\u,\;\nabla p\in L^t(\complement C_{R_0}) ,\quad\forall\,
t>1. 
\end{array}
\end{equation} 

We can say just a little bit more about the second derivatives of $p$. Assuming for simplicity $\u={\bf 0}$ on $\partial\Omega$,
the solution $p$  of the equation
$$
\Delta p+\hbox{\rm div}\,(\u\cdot\nabla\u)=0
$$
can be written
$$
p(x)=-{1\over2\pi}\int_{{\Bbb R}^2}(\log|x-y|)\hbox{\rm div}\,(\u\cdot\nabla\u)(y){\rm d a}_y+\varpi(x)= Q(x)+\varpi(x), 
$$
where  $\varpi(x)$ is a simple layer harmonic potential with a density having zero integral mean over $\partial\Omega$. By
(\ref{Poklmj}) and Theorem II.2 of   \cite{CLMS} $\hbox{\rm div}\,(\u\cdot\nabla\u)\in{\cal H}^t$, for all $t>2/3$. Therefore, by
well--known results about singular integrals (see, $e.g$, \cite{Stein} p. 136)  we have that 
$$
\nabla_2 p(x)=\nabla_2 Q(x)+ O(r^{-3}),
$$ 
with $\nabla_2 Q(x)\in {\cal H}^t$, for all $t>2/3$.\hfill$\diamond$
\end{Remark}

\begin{Remark}
\label{Hamelce}
It is worth noting that   (\ref{hg2jgh0}) is sharp in the sense that, in general, it cannot be replaced by 
\begin{equation}
 \label{lllmnj8}
\nabla p=O(r^{-1-\epsilon}), 
\end{equation}
for some positive $\epsilon$. 
 Indeed, the pairs 
\begin{equation}
\label{kakakak}
\begin{array}{l}
\displaystyle u_r={\gamma\over r},\quad u_\theta=\alpha\left({1\over r}-r^{\gamma+1}\right),\\[12pt]
\displaystyle p=-{\gamma^2+\alpha^2\over 2r^2}-{2\alpha^2r^\gamma\over \gamma}+{\alpha^2\over
2\gamma+2}r^{2(\gamma+1)},
\end{array}
\end{equation}
with $\gamma$ and $\alpha$ arbitrary constants, $\gamma+1\ne 0$, define  the {\it Hamel   solutions\/}  (1916) of the
Navier--Stokes equations (see \cite{OAL} p. xi). For $\gamma+1=-\epsilon/4<0$ (\ref{kakakak})  is a $D$--solution which does not satisfy
(\ref{lllmnj8}).

For $\gamma<-1$ (\ref{kakakak}) gives a family of $D$--solutions of the Navier--Stokes problem in $\complement C_1$ with
boundary datum
\begin{equation}
\label{klji89}
u_r=\gamma,\quad  u_\theta=0.
\end{equation}
Therefore, at least for $\gamma+1<0$,    problem (\ref{NSS}) with the condition at infinity 
$$
\lim_{r\to+\infty}\u(x)={\bf 0}
$$  
does not admit a uniqueness theorem in the class of $D$--solutions. 
Let us recall that if $\varphi\in D^{1,q}(\complement C_{R_0})$, $q\in [1,2)$, then there is a constant $\varphi_0$ such that (see
\cite{Galdi1} Lemma II.5.2)
$$
\int_0^{2\pi}|\varphi(r,\theta)-\varphi_0|^q\le {c(q)\over R^{2-q}}  \int_{\complement_{C_{R_0}}}|\nabla\varphi|^q.      
$$
Therefore (\ref{kakakak}) shows that also (\ref{hg2jgh01}) is sharp in the sense that it cannot be replaced by $p\in D^{1,q}(\complement
C_{R_0})$ for some
$q<2$. Moreover, in contrast with (\ref{Poklmj}) a $D$--solution vanishing at infinity and with nonzero outflow cannot belong to
any
$D^{1,q}(\complement C_{R_0})$  for $q<2$. 

Note that  
\begin{equation}
\label{kj˜lkj12}
 \gamma+1<0\:\Rightarrow\;\displaystyle |\gamma| ={1\over 2\pi}\left|\int_{\partial\Omega}\u\cdot\n\right|>1
\end{equation}
so that for $\Omega'=C_1$ the solution  of Theorem \ref{TS}  is not a Hamel solution.\hfill$\diamond$
\end{Remark}
 
\begin{Remark} From Theorem \ref{lkoinb} it follows that
\begin{equation}
\label{hg2jgaa3}
\partial_r\int_0^{2\pi} u_r^2(r,\theta)=O(r^{-1}\log r) 
\end{equation}
and if $\u$ is bounded, then\footnote{By the example in Remark \ref{Hamelce} relation (\ref{hg2jgaa3}) is
sharp if $\u=o(1)$.}
\begin{equation}
\label{hg2jgaa4}
\partial_r\int_0^{2\pi} u_r^2(r,\theta)=o(r^{-1}).
\end{equation}
Indeed, in  the polar coordinate system $(r,\theta)$ (\ref{NSS})$_{1,2}$ read
\begin{equation}
\label{NSSPCS}
\begin{array}{r@{}l}
\displaystyle\partial_r p+ u_r\partial_r u_r+{u_\theta\over r}\partial_\theta u_r-{u_\theta^2\over r}=0\\[10pt]
\displaystyle{1\over r}\partial_\theta p+ u_r\partial_r u_\theta+{u_\theta\over r}\partial_\theta u_\theta+{u_ru_\theta\over r}=0\\[10pt]
\displaystyle{u_r\over r}+\partial_ru_r+{1\over r}\partial_\theta u_\theta=0.
\end{array}
\end{equation} 
Integrating  (\ref{NSSPCS}) over $\theta\in(0,2\pi)$ and taking into account (\ref{NSSPCS})$_3$, we get
\begin{equation}
\label{h11gjhy}
\partial_r\int_0^{2\pi}(p+u_r^2)(r,\theta)={1\over r}\int_0^{2\pi}(u_\theta^2-u_r^2)(r,\theta)
\end{equation}
Hence  (\ref{hg2jgaa3}) follows by (\ref{jhljhu76}) and (\ref{hg2jgh0}).

Multiply  (\ref{NSSPCS}) by $r$ and integrate over $C_R\setminus C_{R_0}$. Then, we have
\begin{equation}
\label{aszxsee44}
{1\over R}\int_{C_R\setminus C_{R_0}}\left({p+u_\theta^2\over r}\right)=\int_0^{2\pi}(p+u_r^2)(R,\theta)-{R_0\over
R}\int_0^{2\pi}(p+u_r^2)(R_0,\theta).
\end{equation}
If $\u$ is bounded, then $\u(r,\theta)$ tends uniformly (in $\theta$) to a constant vector as $r\to+\infty$.  Then (\ref{aszxsee44})
implies that
$$
\lim_{r\to+\infty} u_r^2(r,\theta)=\lim_{r\to+\infty} u_\theta^2(r,\theta)
$$
and (\ref{hg2jgaa4}) folllows from (\ref{h11gjhy}), taking into account  (\ref{hg2jgh0}).\hfill$\diamond$
\end{Remark}

\section {Existence theorems}

We are now in a position to prove our general    existence theorems  of a $D$--solution for problem (\ref{NSS}).

\begin{Theorem}
\label{extdlNSi} 
Let  $\Omega$ be an exterior Lipschitz domain  of $\,{\Bbb R}^2$ 
and let 
\begin{equation}
\label{iposua}
 \a\in L^2(\partial\Omega). 
\end{equation}
  If
\begin{equation}
\label{oooib}
{1\over 2\pi}\left|\int_{\partial\Omega}\a\cdot\n\right|<1,
\end{equation}
 then system $(\ref{NSS})$ has a Leray solution 
$(\u,p)\in D^{1,2}_\delta(\Omega)\times  L^2_{\delta,{\rm loc}}(\overline\Omega)$
such that $(\ref{hgkjhgjgh0})$ holds uniformly and
\begin{equation}
\label{kjhgkh6}
\u=\u_0+o(1),
\end{equation}
 with $\u_0$ constant vector; it satisfies $(\ref{hg2jgh0}) $, $(\ref{hg2jgh2})$ 
  and if $\a$ and/or $\partial\Omega$ are more regular, then so does $(\u,p)$ according to
the regularity results $(\i)-(\i\hbox{\rm v})$ for the solutions of the Stokes problem; in particular, 
  if $\a\in
L^q(\partial\Omega)$ $(q>2)$, then $\u\displaystyle\mathop{\longrightarrow}^{\rm nt}\a$. Moreover, there are positive constants
$\epsilon$ and $\mu_0<1$ depending on $\Omega$ such that
\begin{itemize}
\item [\hbox{\rm (\j)}] if
$\a\in C^{0,\mu}(\partial\Omega)$, then $\u\in C^{0,\mu}_{\rm loc}(\overline\Omega)$ for $\mu\in[0,\mu_0)$;
\end{itemize}
\begin{itemize}
\item  [\hbox{\rm (\j\j)}] if
$\a\in W^{1-1/q,q}(\partial\Omega)$, $q\in(\max\{4/3, 2-\epsilon \},2+\epsilon)$, then $(\u,p)\in W^{1,q}_{\rm
loc}(\overline\Omega)\times L^q_{\rm loc}(\overline\Omega)$; if $\Omega$ is of class $C^1$ we can take $\mu_0=1$ and 
$q\in[4/3,+\infty)$;
\end{itemize}
\begin{itemize}
\item [\hbox{\rm (\j\j\j)}] $\a\in W^{1,2}(\partial\Omega)$, then    
$(\u,p)\in D^{2,2}_\delta(\Omega)\times  D^{1,2}_\delta(\Omega)$.
  
\end{itemize}

\end{Theorem}

\begin{Proof}
We look for  a solution of (\ref{NSS}) in the form $\u=\w+\h$, with $\w\in D^{1,2}_{\sigma,0}(\Omega)$   and
$\h$ defined by (\ref{fgamma}).   As is well--known \cite{BOPI}, \cite{Russo},  \cite{RussoPG},
under assumption (\ref{oooib}) the system 
\begin{equation}
\label{SSFitkim}
\begin{array}{r@{}l}
  \Delta{\w}-({{\bmit{\bmit h}}}+{\w})\cdot\nabla({{{\bmit{\bmit h}}}}+{\w})
+\Delta{\bmit\zeta}   & {}
=\nabla Q\quad \;\;\,
\hbox{\rm in } \Omega_k,\\[4pt]
\hbox{\rm div}\,{\w} & {} =0 \;\; \;\quad\quad\hbox{\rm in } \Omega_{k},
\\[4pt]
{\w} & {} ={\bf 0} \;\;\;\;\quad\;\;
\hbox{\rm on }\partial \Omega_{k}
\end{array}
\end{equation}
(for all $k> k_0> R_0$) has a  solution   
$\w_k\in W^{1,2}_{\sigma,0}(\Omega_{k})$  we extend to all  ${\Bbb R}^2$ by
setting
$\w_k={\bf 0}$ in $\complement\Omega$. Of course, $\w_k$ satisfies the 
equation
\begin{equation}
\label{WS}
 \int_\Omega\nabla\w_k\cdot\nabla{\bmit\varphi} =\int_\Omega(
\h+\w_k)\cdot\nabla{\bmit\varphi}\cdot(\h+\w_k)-\int_\Omega \nabla{\bmit\zeta}\cdot\nabla{\bmit\varphi},
\end{equation}
for all ${\bmit\varphi}\in W^{1,2}_{\sigma,0}(\Omega_{k})$.

Let us show that  if (\ref{oooib}) holds, then there is a positive number
$c_0$ independent of
$k$ such that
\begin{equation}
\label{dadrt}
\int_\Omega|\nabla\w_k|^2\le c_0.
\end{equation}
To prove  (\ref{dadrt})  we use  a well--known reasoning of   J. Leray (see also \cite{BOPI} and \cite{Galdi1} section  VIII.7). 
If (\ref{dadrt}) is not true, then we can find  a sequence  of   solutions  $\{\w_k'\}_{k\in{\Bbb N}}$ 
such that
$$
 \lim_{k\to+\infty}J_k^2=
\lim_{k\to+\infty}\int_\Omega|\nabla
\w_k'|^2=+\infty.
$$
In virtue of (\ref{WS})  the   field
$$
\w_k={\w_k'\over J_k} 
$$
satisfies 
\begin{equation}
\label{DefWSj}
\begin{array}{r@{}l}
\displaystyle{1\over J_k}\int_\Omega\nabla{\bmit\varphi} & {}
\displaystyle\cdot\nabla{\w}_k =\int_\Omega 
{\w}_k \cdot\nabla{\bmit\varphi}\cdot {\w}_k  +{1\over J_k}\int_\Omega 
{\bmit h} \cdot\nabla{\bmit\varphi}\cdot {\w}_k  \\[12pt]
  & {} \displaystyle+{1\over J_k}\int_\Omega 
{\w}_k \cdot\nabla{\bmit\varphi}\cdot {\bmit h}+{1\over J_k^2}\int_\Omega 
({\bmit h} \cdot\nabla{\bmit\varphi}\cdot {\bmit h}-\nabla{\bmit\zeta}\cdot\nabla{\bmit\varphi}),
\end{array} 
\end{equation} 
for all ${\bmit\varphi}\in
W^{1,2}_{0,\sigma}(\Omega_{k})$. Since $\|\nabla{\w}_k \|_{L^2(\Omega)}=1$, by the compactness theorem of F.
Rellich from 
$\{{\w}_k\}_{k\in{\Bbb N}}$   we can extract  a subsequence, we denote by the same symbol, which
converges  strongly in
$L^q_{\rm loc}(\overline\Omega)$, for all
$q\in(1,+\infty)$, and  weakly in $D^{1,2}_0(\Omega)$ to a field ${\w} \in
D^{1,2}_{\sigma,0}(\Omega)$, with  $\|\nabla{\w}\|_{L^2(\Omega)}\le 1$.
Letting $k\to+\infty$ in (\ref{DefWSj}),
 we see that the field ${\w} $ is a weak solution of the  Euler equations 
\begin{equation}
\label{EuE}
\begin{array}{r@{}l}
 {\w}\cdot  \nabla{\w} +\nabla  Q & {} = 0\quad \hbox{\rm
in }\Omega,\\[4pt]
\hbox{\rm div}\,{\w} & {} =0\quad \hbox{\rm
in }\Omega,
\\[4pt]
{\w} & {} ={\bf 0}\quad
\hbox{\rm on }\partial\Omega,
\end{array}
\end{equation}
for some pressure field $ Q\in W^{1,q}_{\rm loc}(\overline\Omega)$, $q\in [1,2)$, constant on
$\partial\Omega$ \cite{KP2}.  
Now, choosing ${\bmit\varphi}=\w_k'$ in 
(\ref{DefWSj}) we get 
\begin{equation}
\label{eqeqdefi}
\begin{array}{r@{}l}
1  & {}\displaystyle=   \int_{\Omega}
{\w}_k\cdot \nabla {\w}_k\cdot{\bmit\sigma}+ \int_{\Omega}
{\w}_k\cdot \nabla {\w}_k\cdot(\h-{\bmit\sigma})\\[12pt]
 & {}\displaystyle +{1\over J_k}\int_\Omega 
({\bmit h} \cdot\nabla{\w}_k\cdot {\bmit h}-\nabla{\w}_k\cdot\nabla{\bmit\zeta}).
\end{array}
\end{equation}
By (\ref{GWtccc})
$$
\left| \int_{\Omega}
{\w}_k\cdot \nabla {\w}_k\cdot{\bmit\sigma}\right|\le  {1\over 2\pi}\left|\int_{\partial\Omega}\a\cdot\n\right|\int_\Omega|\nabla
{\w}_k|^2\le {1\over 2\pi}\left|\int_{\partial\Omega}\a\cdot\n\right|.
$$
Therefore (\ref{eqeqdefi}) yields
\begin{equation}
\label{fdfdf}
\begin{array}{r@{}l}
\displaystyle\ 1- {1\over 2\pi}\left|\int_{\partial\Omega}\a\cdot\n\right|   & {}\displaystyle\le \int_{\Omega}
{\w}_k\cdot \nabla {\w}_k\cdot(\h-{\bmit\sigma})\\[12pt]
 & {}\displaystyle +{1 \over J_k}\int_\Omega 
({\bmit h} \cdot\nabla{\w}_k\cdot {\bmit h}-\nabla{\w}_k\cdot\nabla{\bmit\zeta}).
\end{array}
\end{equation}
Hence, taking into account that by (\ref{srvissima}) 
$$
\begin{array}{l}
 \displaystyle\left|\int_{\Omega}
 {\bmit h} \cdot\nabla{\w}_k\cdot {\bmit h}  
\right| \le  \left\{\int_\Omega |{\bmit h}|^4  \int_\Omega|\nabla {\w}_k|^2 \right\}^{1/2}\le c, \\[12pt]
\displaystyle\left|\int_{\Omega}\nabla{\w}_k\cdot\nabla{\bmit\zeta} 
\right|=\left|\int_{T_{\bar R}}\nabla{\w}_k\cdot\nabla{\bmit\zeta} 
\right| \le \left\{\int_\Omega | \nabla{\w}_k|^2\int_{T_{\bar R}}|\nabla{\bmit\zeta}|^2\right\}^{1/2}\le c  ,  
\end{array}
$$
and letting $k\to+\infty$ in (\ref{fdfdf}), it follows
\begin{equation}
\label {PCINC}
 1-  {1\over 2\pi}\left|\int_{\partial\Omega}\a\cdot\n\right| \le\int_{\Omega}
{\w} \cdot\nabla {\w} \cdot{\bmit\zeta}.
\end{equation}
Taking into account that  $ Q$   is   constant  on $\partial\Omega$ (say $ Q_0$) and  ${\bmit\zeta}$ is divergence free
in
${\Bbb R}^2$ 
 we have
\begin{equation}
\label {SCINC}
 \int_{\Omega}
{\w} \cdot\nabla {\w} \cdot{\bmit\zeta}
=-\int_\Omega{\bmit\zeta}\cdot\nabla
 Q=- Q_0\int_{\partial\Omega}{\bmit\zeta}\cdot \n=0.
\end{equation}
Since, under assumption (\ref{oooib}), (\ref{PCINC}) and (\ref{SCINC}) are incompatible, we conclude that
(\ref{dadrt}) is true. Therefore, by the compactness theorem of F. Rellich from
$\{\w_k\}_{k\in{\Bbb N}}$ we can extract a subsequence which converges strongly in $L^q_{\rm loc}(\overline\Omega)$ and
weakly  in $D^{1,2}(\Omega)$ to a field $\w\in D^{2,1}_{\sigma,0}(\Omega)$ that a well--known argument shows to be a   solution
of equations (\ref{SSFitkim}) (see, $e.g.$, \cite{Sohr}  Ch. 5). 

 (\ref{kjhgkh6})
is proved in \cite{GaldiDD}, \cite{GWe},  while (\ref{hgkjhgjgh0}),    (\ref{hg2jgh0}), (\ref{hg2jgh2}) 
are consequence of the fact that $\u$ is a $D$--solution. As far as the boundary datum is concerned, let us note that $\u={\bmit h}+\w$
attains
$\a$ in the following sense
$$
{\bmit h}\displaystyle\mathop{\longrightarrow}^{\rm nt}\a,\quad \hbox{\rm tr}_{\vert\partial\Omega}\w={\bf
0},
$$
where $\hbox{\rm tr}_{\vert\partial\Omega}$ stands for the trace operator in the Sobolev space
$D^{1,2}_0(\Omega)$. If
$\a\in L^q(\partial\Omega)$
$(q>2)$ then ${\bmit h}\in L^{2q}(\Omega)$ so that by well--known estimates about solution of the Stokes problem
$\w\in W^{1,s}_{\rm loc}(\overline\Omega)$, for some  $s>2$. Hence by Sobolev's lemma it follows that $\w$ is continuous in
$\overline\Omega$ and ${\bmit u}\displaystyle\mathop{\longrightarrow}^{\rm nt}\a$. Of course, if $\a\in C(\partial\Omega)$, then $\u\in
C^\infty(\Omega)\cap C(\overline\Omega)$.
Moreover, (\j)--(\j\j\j) are consequence of (\i)--(\i v),
\end{Proof}
 
It is not difficult to see that the above argument can be repeated for boundary data $\a\in W^{-1/q,q}(\partial\Omega)$, $q\ge 4$,
provided we make use of the divergence free extension of $\a$ defined in Lemma \ref{L01} and assume that\footnote{By
$\langle\a,\n\rangle $ we mean the value of  the functional $\a\in W^{-1/q,q}(\partial\Omega)$ at $\n$.}
\begin{equation}
\label{jljaakh7}
 \left|\langle\a,\n\rangle \right|<2\pi.
\end{equation}
Indeed, the following theorem holds.
\begin{Theorem}
\label{extdlNSi2}
Let  $\Omega$ be an exterior   domain  of $\,{\Bbb R}^2$ of class $C^{1,1}$. If $\a\in W^{-1/q,q}(\partial\Omega)$, $q\ge 4$, satisfies
$(\ref{jljaakh7})$, then
 $(\ref{NSS})$ has a D--solution  
$$
\u\in L^q_{\rm loc}(\overline\Omega) \cap L^\infty(\complement C_{R_0}).
$$
Moreover, 
\begin{itemize}
\item if $\a\in C^{1,\mu}(\partial\Omega)$, $\mu\in(0,1)$, then  
$$
(\u,p)\in C^{1,\mu}_{\rm loc}(\overline\Omega)\times C^{0,\mu}_{\rm loc}(\overline\Omega),
$$
\item if $\Omega$ is of class $C^k$ $(k\ge 2)$  and $\a\in W^{k-1/q,q}(\partial\Omega)$, then   
$$
(\u,p)\in W^{k,q}_{\rm loc}(\overline\Omega)\times W^{k-1,q}_{\rm loc}(\overline\Omega).
$$
\end{itemize}
\end{Theorem}

\bigskip

Let $\Omega$ be polar symmetric, $i.e$,
$$
(x_1,x_2)\in\Omega\Rightarrow (-x_1,-x_2)\in\Omega.
$$
If $\a$ is polar symmetric, $i.e$,
\begin{equation}
\label{jgfjhgf5}
\a(\zeta)=-\a(-\zeta),
\end{equation}
for all $\zeta\in\partial\Omega$, then the field $\h$ can be constructed polar symmetric  and we can find a polar symmetric
solution of (\ref{SSFitkim}). As a consequence,  the solution $(\u,p)$ in Theorem \ref{extdlNSi} satisfies the symmetry properties
\begin{equation}
\label{SPr}
\begin{array}{r@{}l}
\u(x)  & {} =-\u(-x) ,\\[4pt]
 p(x)& {}=p(-x),
\end{array}
\end{equation}
for all $x\in\Omega$.  Since by (\ref{SPr})$_1$
$$
\int_0^{2\pi}\u(R,\theta)={\bf 0},
$$
for all $R>R_0$, by Poincar\'e's inequality we get
\begin{equation}
\label{jkiu89}
\int_{T_R}|\u|^2\le cR^2\int_{T_R}|\nabla\u|^2,
\end{equation}
with $c$ independent of $R$. Therefore, by the trace theorem and (\ref{jkiu89}) 
\begin{equation}
\label{hjuddf5}
\int_0^{2\pi}|\u|^2(R,\theta)\le c\left\{{1\over R^2}\int_{T_R}|\u|^2+\int_{T_R}|\nabla\u|^2\right\}\le c\int_{\complement
C_R}|\nabla\u|^2,
\end{equation}
with $c$ independent of $R$. Hence it follows
\begin{equation}
\label{hjyu78}
\lim_{R\to+\infty}\int_0^{2\pi}|\u|^2(R,\theta)=0.
\end{equation}
By virtue of the results of \cite{GaldiDD}, \cite{GW}, (\ref{hjyu78}) is sufficient to conclude that
\begin{equation}
\label{mbvmbvn7}
\lim_{r\to+\infty}\u(r,\theta)={\bf 0},
\end{equation}
uniformly in $\theta$. Therefore we can state
\begin{Theorem}
\label{TS}
Let  $\Omega$ be a polar symmetric  exterior Lipschitz domain  of $\,{\Bbb R}^2$. If $\a\in L^2(\partial\Omega)$ is polar symmetric
and satisfies
$(\ref{oooib})$,
then $(\ref{NSS})$, $(\ref{mbvmbvn7})$  has a Leray  solution which satisfies \hbox{\rm (\j)}--\hbox{\rm (\j\j\j)}
and $(\ref{hgkjhgjgh0})$,
$(\ref{kjhgkh6})$, $(\ref{hg2jgh0})$, $(\ref{hg2jgh2})$. If $\Omega$ is of class $C^{1,1}$, then we can assume $\a\in
W^{-1/4,4}(\partial\Omega)$.
\end{Theorem}

It is evident that (\ref{hjuddf5}) holds for every polar symmetric $D$--solution.
Hence it follows
\begin{Theorem}
\label{ggbhy} 
A polar symmetric D--solution tends to zero at infinity.
\end{Theorem}
 
The Hamel solutions (\ref{kakakak}) are polar symmetric and for $\gamma<-1$ have finite Dirichlet integrals. Since we can
choose $\gamma$ close to $-1$ as we want, we see that Theorem \ref{ggbhy} is sharp in the sense that (at least for $\gamma<-1)$ a polar
symmetric solution cannot tend  to zero at infinity as $r^{-\epsilon}$ for some positive $\epsilon$. Note that by virtue of
(\ref{kj˜lkj12}) these considerations do not apply to the $D$--solution   of Theorem \ref{TS}.

\begin{Remark} 
Existence of a solution of  (\ref{NSS}) with less regular boundary data (say in $L^q(\partial\Omega)$ and $W^{-1/q,q}(\partial\Omega)$)
have been studied by several authors  for bounded and regular  domains with connected boundaries (see
\cite{Amrouche}
\cite{GSS}, \cite{Giga},  \cite{RussoPG} and the references therein). As far as Lipschitz domains are concerned, to the best of
our knowledge  problem  (\ref{NSS}) (with
$L^q(\partial\Omega)$ data) has  been considered  only for bounded domains in 
 \cite{ARAnnali},  \cite{Russo},  \cite{RussoPG} under a restriction on the flux, in \cite{DM} for small data and in \cite{RStrara}
for   domains symmetric with respect to the $x_1$ axis, $a_1$  pair function of $x_2$ and $a_2$ odd function of $x_2$. In
\cite{ARSub} the classical Finn--Smith theorem \cite{FS} has been proved for Lipschitz domains and boundary data in
$L^\infty(\partial\Omega)$. \hfill$\diamond$
\end{Remark}
 
\bigskip

As we said in the introduction, there is another technique, based on a Galerkin's type scheme and due to H. Fujita \cite{Fu}  to
prove existence of a $D$--solution of (\ref{NSS}), we shall call {\it Fujita solution\/}. It reduces the problem to find the 
uniform estimate
$$
\int_\Omega|\nabla\w|^2\le c,
$$
for every  solution  $\w\in D^{1,2}_\sigma(\Omega)$ with compact support in $\Omega$ of the system\footnote{Clear
expositions of this approach can be find in   \cite{Galdi1}, \cite{OAL}, \cite{Temam}.}
\begin{equation}
\label{Ssskim}
\begin{array}{r@{}l}
  \Delta{\w}-({{\bmit{\bmit h}}}+{\w})\cdot\nabla({{{\bmit{\bmit h}}}}+{\w})
+\Delta{\bmit\zeta}   & {}
=\nabla Q\quad  
\hbox{\rm in } \Omega ,\\[4pt]
\hbox{\rm div}\,{\w} & {} =0 \;\; \;\;\quad \hbox{\rm in } \Omega,
\\[4pt]
{\w} & {} ={\bf 0}\;\; \;\,\quad
\hbox{\rm on }\partial \Omega,
\end{array}
\end{equation} 
where now  $\h$ has the form (\ref{fgamma}) with ${\bmit\zeta}=\nabla^\perp(g_{\delta_0}{\eta})$,
${\bmit v}=\nabla^\perp{\eta}$ and
$g_{\delta_0}$ Leray--Hopf cut--off function  of the regularized distance $\varrho(x)$   equal to $1$ for $\varrho(x)\le c_1\delta_0$
and  vanishing for 
$\varrho(x)\ge c_2\delta_0$. A
straightforward calculation yields the relation 
$$
\int_\Omega|\nabla\w|^2\le
\int_\Omega\w\cdot\nabla\w\cdot\h+c(\delta_0)\left\{\int_{\partial\Omega}|\a|^2+\left[\int_{\partial\Omega}|\a|^2\right]^2\right\}
$$
By  a classical procedure we have (see, $e.g.$, \cite{Galdi1}, \cite{Temam})
$$
\left|\int_\Omega\w\cdot\nabla\w\cdot{\bmit\zeta}\right|\le \alpha(\delta_0)\int_\Omega|\nabla\w|^2,
$$
with
$$
\lim_{\delta_0\to 0}\alpha(\delta_0)=0.
$$
Moreover, by (\ref{GWtccc})
$$
\left|\int_\Omega\w\cdot\nabla\w\cdot{\e_r\over r}\right|\le \int_\Omega|\nabla\w|^2.
$$
so that if (\ref{oooib}) holds, then  there is a constant $c$ independent of $\w$ such that 
\begin{equation}
\int_\Omega|\nabla\w|^2\le c\left\{\int_{\partial\Omega}|\a|^2+\left[\int_{\partial\Omega}|\a|^2\right]^2\right\}.
\end{equation} 
Therefore, taking also into account (\ref{jknhhhu78}), we have
\begin{Theorem}
Let  $\Omega$ be an exterior Lipschitz domain  of $\,{\Bbb R}^2$. If $\a$ satisfies $(\ref{iposua})$ and  $(\ref{oooib})$
then  $(\ref{NSS})$ has a Fujita solution $(\u,p)$ such that
\begin{equation}
\label{kkklioj}
\int_\Omega \delta |\nabla\u|^2 \le
c\left\{\int_{\partial\Omega}|\a|^2+\left[\int_{\partial\Omega}|\a|^2\right]^2\right\}.
\end{equation}
\end{Theorem}

\begin{Theorem}
\label{hjlhy00}
Let  $\Omega$ be an exterior   domain  of $\,{\Bbb R}^2$ of class $C^{1,1}$. If $\a\in W^{-1/4,4}(\partial\Omega)$, 
satisfies
$(\ref{jljaakh7})$, then
 $(\ref{NSS})$ has a D--solution $\u\in L^4_{\rm loc}(\overline\Omega)$. 
\end{Theorem}

It is quite evident that the Fujita solutions  enjoys all the regularity properties as those of the Leray solution.
The only substantial difference is that the latter is always bounded while by Theorem \ref{Ami1} we know that  the
former is bounded   for zero outflow. 

\begin{Remark}
\label{DDMMP}
It is not difficult to see that Theorems \ref{Ami1} -- \ref{hjlhy00} can be stated for the system  
\begin{equation}
\label{NSSNO}
\begin{array}{r@{}l}
 \Delta{\u}- \u\cdot \nabla{\u} - \nabla p  & {} =\f \quad \hbox{\rm
in }\Omega,\\[4pt]
\hbox{\rm div}\,{\u} & {} =0\,\,\quad \hbox{\rm
in }\Omega,
\\[4pt]
{\u} & {} ={\a} \,\quad
\hbox{\rm on }\partial\Omega 
\end{array}
\end{equation}
in the more general exterior domain\footnote{  Lemma  \ref{L0} continues to hold for the domain (\ref{L0gd}); in such a case 
$$
{\bmit\sigma}=-{1\over  2\pi}\sum_{i=1}^m {( x-x_i)\over|x-x_i|^2 } \int_{\partial\Omega_i}\a\cdot\n,
$$
where $x_i$ is a fixed point of $\Omega_i$.}
\begin{equation}
\label{L0gd}
\Omega={\Bbb
R}^2\setminus\overline{\Omega'},\quad\Omega'=\bigcup_{i=1}^m\Omega_i,\quad
\overline\Omega_i\cap\overline\Omega_j=\varnothing,\;i\ne j,
\end{equation} 
with $\partial\Omega_i$ Lipschitz and connected, provided
\begin{equation}
\label{yuijhj87}
\f\in{\cal H}^1(\Omega)
\end{equation}
vanishes outside a bounded set\footnote{This is not necessary for the existence of a $D$--solution
 to (\ref{NSSNO}). It is worthy to note that for the validity of Theorem \ref{Ami1} it is sufficient that
(\ref{ioupou7}) holds for a circumference surrounding $\overline{\Omega'}$.}, is polar symmetric in Theorems
\ref{TS},
\ref{ggbhy} and
$$
{1\over 2\pi}\sum_{i=1}^m  \left|\int_{\partial\Omega_i}\a\cdot\n\right|<1 
$$
($\sum_{i=1}^m |\langle\a,\n\rangle|<2\pi$ for $\Omega$ of class $C^{1,1}$ and $\a\in
W^{-1/4,4}(\partial\Omega)$). Under assumption (\ref{yuijhj87})  
$(\u,p)$ satisfies  (\ref{NSSNO}) almost everywhere in
$\Omega$ and $\u$ is continuous in
$\Omega$ \cite{Catafalco}. Moreover, if $\hbox{\rm div}\,\f\in {\cal H}^1(\Omega)$, then $p$ is continuous in $\Omega$. Moreover,
(\ref{kkklioj}) becomes
$$
\|\u\|_{D^{1,2}_\delta(\Omega)} \le
c\left\{\|\a\|_{L^2(\partial\Omega)}+\|\f\|_{{\cal H}^1}+\big[\|\a\|_{L^2(\partial\Omega)}+\|\f\|_{{\cal
H}^1}\big]^2\right\}.\eqno\diamond
$$
\end{Remark}

\section{A uniqueness theorem}
 Uniqueness of a $D$--solution converging to a nonzero vector at infinity\footnote{Recall that uniqueness does not hold when the
$D$--solution is zero at infinity  at least for large Reynolds numbers (see Remark \ref{Hamelce}).} is a complicated question and 
only in few cases we know as to determine  small uniqueness classes (see
\cite{FS} and \cite{Galdi1} Ch. X). We aim at observing now as uniqueness could be linked with the boundary data at
least in particular situations: {\it the potential flows\/}. 

Let us consider the harmonic simple layer potential  with density
$\psi$\footnote{It can be proved that every harmonic function $u$ in a Lipschitz exterior domain  $\Omega$ of ${\Bbb R}^2$ such
that $u=o(r)$ and $\hbox{\rm tr}_{\vert\partial\Omega} u\in  W^{1,2}(\partial\Omega)$, is expressed by  (\ref{SLPP})
for some $\psi\in L^2(\partial\Omega)$.}
\begin{equation}
\label{SLPP}
v(x) ={1\over 2\pi}\int_{\partial\Omega}\psi(\zeta)\log|x-\zeta|{\rm d} s_\zeta
\end{equation}
and the Navier--Stokes problem 
\begin{equation}
\label{NSSarmo}
\begin{array}{r@{}l}
 \Delta{\u}- \u\cdot \nabla{\u}  & {} = \nabla p\quad  \hbox{\rm
in }\Omega,\\[4pt]
\hbox{\rm div}\,{\u} & {} =0 \! \qquad \hbox{\rm
in }\Omega,
\\[4pt]
{\u} & {} ={\a}\!\!\qquad
\hbox{\rm on }\partial\Omega,\\[4pt]
\u & {} =\e_1+o(1),
\end{array}
\end{equation}
with the boundary datum 
\begin{equation}
\label{BDA}
\a(\xi)= \nabla v(\xi)+\e_1.
\end{equation}
Note that by 
$$
\int_{\partial\Omega}\a\cdot\n=\int_{\partial\Omega}\partial_n v=\int_{\partial\Omega}\psi.
$$
 
The pair 
\begin{equation}
\label{hhjjucvbn6}
(\nabla v+\e_1,-{\textstyle{1\over 2}}|\nabla v|^2-\partial_1 v)
\end{equation}
is a $D$--solution to (\ref{NSSarmo}), (\ref{BDA}). By what we said above, we could have   other {\it a priori different\/}
$D$--solutions  of (\ref{NSSarmo})--(\ref{BDA}), as the Finn--Smith, Galdi,   Leray and   Fujita solutions.  Let shows that
if 
\begin{equation}
\label{fluipoi}
\int_{\partial\Omega}|\psi|<2\pi,
\end{equation}
then all these solutions coincide. Indeed, the following theorem holds.
\begin{Theorem} 
Let $\Omega$ be an exterior Lipschitz domain of ${\Bbb R}^2$.  If $\psi\in L^2(\partial\Omega)$ satisfies $(\ref{fluipoi})$,
then $(\ref{hhjjucvbn6})$ is unique in the class of all $D$--solutions.
\end{Theorem}

\begin{Proof}
Let $(\u+\w,p+ Q)$ be another $D$--solution to (\ref{NSS}), (\ref{BDA}). Then $(\w,Q)$
 satisfies the equation
\begin{equation}
\label{NSSuniq}
\begin{array}{r@{}l}
 \Delta{\w}- (\u+\w)\cdot \nabla{\w}-\w\cdot\nabla\u  & {} = \nabla Q\quad  \hbox{\rm
in }\Omega,\\[4pt]
\hbox{\rm div}\,{\w} & {} =0 \! \qquad \hbox{\rm
in }\Omega,
\\[4pt]
{\w} & {} ={\bf 0}\!\!\qquad
\hbox{\rm on }\partial\Omega,\\[4pt]
{\w} & {} =o(1).
\end{array}
\end{equation}
Let $g(r)$ be a regular function, equal to 1 in $C_R$, vanishing outside $C_{2R}$ and such that $|\nabla g|\le cR^{-1}$. Then by a
standard computation we get 
\begin{equation}
\label{gggt65}
\begin{array}{r@{}l}
\displaystyle\int_\Omega g|\nabla \w|^2 & {} \displaystyle = \int_{T_R}\big[{\textstyle{1\over 2}}|\w|^2(\u+\w)+(\u\cdot
\w+Q)\w\big]\cdot\nabla g\\[12pt]
& {} \displaystyle+\int_\Omega
 g \w\cdot\nabla\w\cdot\u.
\end{array}
\end{equation}
By H\"older's inequality and (\ref{Poklmj})
$$
\begin{array}{l}
 \displaystyle\left|\int_{\Omega}
|\w|^2( \u -\e_1)\cdot\nabla g
\right| \le {c\over
R}\left\{\int_{T_R}|\u-\e_1|^4\right\}^{1/4}\left\{\int_{T_R}|\w|^8\right\}^{1/4}\left\{\int_{T_R}\right\}^{1/2} =o(1),
\\[12pt]
 \displaystyle\left|\int_{\Omega}
|\w|^2\partial_1 g
\right| \le {c\over
R} \left\{\int_{T_R}|\w|^4\right\}^{1/2}\left\{\int_{T_R}\right\}^{1/2} =o(1),
\\[12pt]
 \displaystyle\left|\int_{\Omega}
|\w|^2\w\cdot\nabla g
\right| \le {c\over
R}\left\{\int_{T_R}|\w|^6\right\}^{1/3} \left\{\int_{T_R}\right\}^{1/2} =o(1),
\\[12pt]
\displaystyle\left|\int_{\Omega}
Q\w \cdot\nabla g
\right| \le {c \over R}\left\{\int_{T_R}|\w|^4\int_{T_R}|Q|^4\right\}^{1/4}\left\{ 
\int_{T_R}\right\}^{1/2} =o(1).   
\end{array}
$$
Likewise,
$$
\left|\int_{\Omega}(\u\cdot\w) \w \cdot\nabla g\right|  =o(1).
$$
Moreover,
$$
\left|\int_\Omega g\w\cdot\nabla\w_1\right|=\left|\int_{T_R} w_1\w\cdot\nabla g\right|\le
{c\over
R} \left\{\int_{T_R}|\w|^4\right\}^{1/2}\left\{\int_{T_R}\right\}^{1/2} =o(1)
$$ 
By (\ref{GWtccc}) 
$$
\begin{array}{r@{}l}
\displaystyle\left|\int_\Omega \w\cdot\right. & {}\displaystyle\left.\nabla\w\cdot\nabla v \right|= \left| \int_\Omega 
v \nabla\w\cdot\nabla\w^{\rm T}\right|\\[10pt]
 & {}  \displaystyle  ={1\over2\pi}\left|\int_{\partial\Omega}\psi(\zeta)\int_\Omega \nabla\w\cdot\nabla\w^{\rm
T}\log|x-\zeta|{\rm d}a_x \right|\\[10pt]
  & {} \displaystyle
={1\over2\pi}\left|\int_{\partial\Omega}\psi(\zeta)\int_\Omega\w\cdot\nabla\w\cdot{(x-\zeta)\over|x-\zeta|^2}{\rm d}a_x\right|\le 
{\|\psi\|_{L^1(\partial\Omega)} \over2\pi}\int_\Omega|\nabla\w|^2  .
\end{array}
$$
for all $\w\in D^{1,2}_{\sigma,0}(\Omega)$.
Therefore, letting $R\to+\infty$ in  (\ref{gggt65}), we have 
$$
\big(2\pi-\|\psi\|_{L^1(\partial\Omega)}\big)\int_\Omega|\nabla\w|^2\le 0.
$$
Hence uniqueness follows at once.
\end{Proof}

\begin{Remark}  Note that if 
$$
v(x)= {\mu\log r\over2\pi}, 
$$
then
$$
\int_{\partial\Omega}\a\cdot\n=\mu
$$
and (\ref{fluipoi}) takes the weaker form.
$$
|\mu|<2\pi.\eqno\diamond
$$
\end{Remark}

\begin{Remark} 
\label{AMIP} 
When $\partial\Omega$ is connected and $\a={\bf 0}$, $\u_0=\e_1$, a solution of the equations
\begin{equation}
\label{NSSde}
\begin{array}{r@{}l}
 \Delta{\u}- \u\cdot \nabla{\u}  & {} = \nabla p\quad  \hbox{\rm
in }\Omega,\\[4pt]
\hbox{\rm div}\,{\u} & {} =0 \! \qquad \hbox{\rm
in }\Omega,
\\[4pt]
{\u} & {} ={\bf 0}\!\!\qquad
\hbox{\rm on }\partial\Omega,\\[4pt]
\displaystyle\lim_{r\to+\infty}\u(x) & {}=\e_1
\end{array}
\end{equation}
represents the translational  motion (with velocity $-\e_1$) of an object in a Navier--Stokes fluid assumed to be at rest at
infinity. As we remarked in this paper, problem (\ref{NSSde}) is completely open. By the Leray argument we know that the sequence
of solutions of the systems
\begin{equation}
\label{NSSdeL}
\begin{array}{r@{}l}
 \Delta{\u}_k- \u_k\cdot \nabla{\u}_k  & {} = \nabla p_k\quad  \hbox{\rm
in }\Omega_k,\\[4pt]
\hbox{\rm div}\,{\u}_k & {} =0\; \qquad \hbox{\rm in }\Omega_k,
\\[4pt]
{\u}_k & {} ={\bf 0}\,\qquad
\hbox{\rm on }\partial\Omega_k,\\[4pt]
 \u_k & {}=\e_1\!\qquad
\hbox{\rm on }\partial C_k
\end{array}
\end{equation}
converges to a $D$--solution to $(\ref{NSSde})_{1,2,3}$ and there is a constant vector $\u_0$ such that \cite{GWe} 
$$
\lim_{r\to+\infty}\u(r,\theta)=\u_0,
$$
uniformly on $\theta$. However, we do not know $\u_0$ so that   in principle  it could be zero and the Leray construction
could even yield
 the trivial solution, as it happens for the Stokes paradox (see Section \ref{SPARA}). C.J. Amick excluded this possibility for
domains of class
$C^3$, symmetric with respect to the $x_1$--axis \cite{Amick2} (see also \cite{GaldiDD}). This result has been recently extended
to symmetric Lipschitz   domains in \cite{RussoARM}. 
\end{Remark}

\section{Some remarks on the Stokes paradox}
\label{SPARA}
More in general, introducing the Reynolds number $\lambda=vl/\nu$, with $v$, $l$ reference  velocity and reference length, and
$\nu$ kinematical viscosity of the fluid, the steady--state Navier--Stokes problem in an exterior Lipschitz domain $\Omega$ of
${\Bbb R}^2$ writes
\begin{equation}
\label{NSSLa}
\begin{array}{r@{}l}
 \Delta{\u}- \lambda\u\cdot \nabla{\u}-\nabla p & {} = \f\quad  \hbox{\rm
in }\Omega,\\[4pt]
\hbox{\rm div}\,{\u} & {} =0\,\quad \hbox{\rm
in }\Omega,
\\[4pt]
{\u} & {} ={\a}\quad
\hbox{\rm on }\partial\Omega,\\[4pt]
\displaystyle\lim_{r\to+\infty}\u(x)  & {} ={\u}_0.
\end{array}
\end{equation}
Of course, for $\lambda=0$ (\ref{NSSLa}) reduces to the Stokes problem
\begin{equation}
\label{SSLa}
\begin{array}{r@{}l}
 \Delta{\u} -\nabla p & {} = \f\quad  \hbox{\rm
in }\Omega,\\[4pt]
\hbox{\rm div}\,{\u} & {} =0\,\quad \hbox{\rm
in }\Omega,
\\[4pt]
{\u} & {} ={\a}\quad
\hbox{\rm on }\partial\Omega,\\[4pt]
\displaystyle\lim_{r\to+\infty}\u(x)  & {} ={\u}_0.
\end{array}
\end{equation}

In this section we aim at comparing the known results  for systems (\ref{NSSLa}), (\ref{SSLa}). 
It is well--known  a $D$--solution of (\ref{SSLa})$_{1,2,3}$ exists  and converges to a constant vector, but
contrary to what happens in the nonlinear case, we know   that  (\ref{SSLa})  has a solution
 if and only if $\a\in L^2(\partial\Omega)$, $\f\in{\cal H}^1$ and $\u_0$ satisfy the compatibility condition \cite{GS},
\cite{RussoPG}
\begin{equation}
\label{CC}
\int_{\partial\Omega}(\a-\u_0)\cdot{\bmit T}(\h_i,p_i)\cdot\n+\int_\Omega\f\cdot\h_i=0, \quad i=1,2,
\end{equation}
with $(\h_i,p_i)$ solution  of\footnote{The solutions of (\ref{nnnmuij}) span a linear space of dimension two and every $\h$ behaves at
infinity as $\log r$} 
\begin{equation}
\label{nnnmuij}
\begin{array}{r@{}l}
 \Delta{\h}_i & {} = \nabla p_i\quad  \hbox{\rm
in }\Omega,\\[4pt]
\hbox{\rm div}\,{\h}_i & {} =0 \; \qquad \hbox{\rm
in }\Omega,
\\[4pt]
{\h}_i & {} ={\e}_i\;\;\;\quad
\hbox{\rm on }\partial\Omega,\\[4pt]
\displaystyle\h_i & {} =o(r).
\end{array}
\end{equation}
 For instance, if $\partial\Omega$ is an ellipse, then $[{\bmit T}(\h_i,p_i)\cdot\n](\xi)=({\bmit\xi}\cdot\n(\xi))\e_i$
\cite{MRS} and for $\f={\bf 0}$ (\ref{CC}) writes
$$
\int_{\partial\Omega}(\a-\u_0)_i({\bmit\xi}\cdot\n(\xi))=0,\quad i=1,2.
$$
In particular, if $\u$ is a $D$--solution of (\ref{SSLa})$_{1,2,3}$, then for large $R$ 
$$
{1\over 2\pi}\int_0^{2\pi}u_i(R,\theta){\rm d}\theta 
$$
is constant and gives the vector to which $\u$ tends at infinity (Picone's mean theorem at infinity).

Since
$$
\int_{\partial\Omega}{\bmit T}(\h_i,p_i)\cdot\n\ne{\bf 0},
$$
from (\ref{CC}) it follows that  if $\a=\f={\bf 0}$, then the only solution to (\ref{SSLa}) is the trivial one  so that
necessarily
$\u_0$ must be zero (Stokes' paradox).
The results of  R. Finn \& D.R. Smith \cite{FS} (see also \cite{Galdi1}, \cite{ARSub}) and C.J. Amick \cite{Amick2} (see Remark
\ref{AMIP}) allow  us to state
\begin{Theorem} 
Let $\Omega$ be an exterior Lipschitz domain of ${\Bbb R}^2$. If $\lambda$ is sufficiently small or $\Omega$ is
symmetric with respect to an axis, then the Stokes paradox holds if and only if $\lambda=0$.
\end{Theorem}

Of course, a $D$--solution of (\ref{NSSLa}) must satisfy (\ref{CC}) whenever the integrals
$$
\int_\Omega \u\cdot \nabla\u\cdot\h
$$
make  sense. In particular, taking into account that if $\Omega$  is polar symmetric, then $\h(x)=\h(-x)$ for all
$x\in\Omega$, we see that the solution of Theorem  \ref{TS} satisfy (\ref{CC}).

\end{document}